%% file: paper.tex
\documentclass[runningheads]{llncs}
\usepackage[utf8]{inputenc}
\usepackage{times}
\usepackage{helvet}
\usepackage{courier}

\usepackage{xspace}
\usepackage{graphicx}
\usepackage{subfigure}
\usepackage{amsmath}
\usepackage{amssymb}
\usepackage{latexsym}
\usepackage{multirow}
\usepackage{array}
\usepackage{url}
\usepackage{paralist}
\usepackage{placeins}
\usepackage{cite}
\graphicspath{{figs/}}
\input{macros}
\usepackage[utf8]{inputenc}
\usepackage[lined,boxed,commentsnumbered]{algorithm2e}

\begin{document}
\mainmatter
\title{Link Prediction and the Role of Stronger Ties in Networks of Face-to-Face Proximity}



\author{Christoph Scholz \and Martin Atzmueller \and Gerd Stumme}

\institute{Knowledge \& Data Engineering Group, University of Kassel, \\
34121 Kassel, Germany\\
\email{\{scholz,atzmueller,stumme\}@cs.uni-kassel.de}
}

\maketitle

\begin{abstract}
Understanding the structures why links are formed is an important and prominent research topic. In this paper, we therefore consider the link prediction problem in face-to-face contact networks, and analyze the predictability of new and recurring links. Furthermore, we study additional influence factors, and the role of stronger ties in these networks. Specifically, we compare neighborhood-based and path-based network proximity measures in a threshold-based analysis for capturing temporal dynamics. 
The results and insights of the analysis are a first step onto predictability applications for human contact networks, for example, for improving recommendations.
\end{abstract}






\input{introduction}
\input{relatedwork}
\input{rfid-dataset}

\input{measures}
\input{analysis}

\input{conclusions}

\section*{Acknowledgements}

This work has been supported by the VENUS research cluster at the interdisciplinary Research Center for Information System Design (ITeG) at Kassel University.
We utilized active RFID technology which was developed within the SocioPatterns project, whose generous support we kindly acknowledge. 
Our particular thanks go the SocioPatterns team, especially to Ciro Cattuto, who enabled access to the Sociopatterns technology, 
and who supported us with valuable information concerning the setup of the RFID technology.


\bibliographystyle{abbrv}

\end{document}

%% file: macros.tex
\newcommand{\eg}{e.\,g.,\xspace}

\newcommand{\ie}{i.\,e.,\xspace}

\newcommand{\conferator}{\textsc{Conferator}\xspace}

\newcommand{\furl}[1]{\footnote{\url{#1}}}

%% file: introduction.tex
\section{Introduction}\label{sec:introduction}

Link prediction is a powerful approach in various application contexts, for example, for supporting recommendation systems. Typically, it leverages \emph{crowd} or \emph{collective} intelligence~\cite{MLD:10,Leimeister:10} captured by a set of actions or measurements; from these, social interaction networks~\cite{MABHS:11,51seitenPaper,MASH:13} between different actors are then derived, and investigated using different link prediction measures. The analysis of such measures can then help in order to understand structural mechanisms of link creation, its dynamics, and for building effective link prediction algorithms. With the growing amount of social data, ubiquitous systems, and mobile social media applications for participatory sensing, link analysis is receiving increased attention. This especially relates to the dynamics of link creation~\cite{AdaAda03}, \eg concerning their mobility~\cite{Wang2011,barabasi2002} and dynamic behavior~\cite{watts1998smallworld,TB:09}.

This paper focuses on the analysis of social networks captured in mobile and ubiquitous settings -- utilizing data from offline networks captured by RFID tags.
In particular, we analyze the predictability of links in such networks of face-to-face proximity comparing several path-based and neighborhood-based network proximity measures.
 While there is a large body of research concerning \emph{online} social networks, \eg~\cite{Kleinberg2003,Murata2007,Hui2005,Lu2009,Katz53,zhou2009,Leskovec2008,Leskovec2010}, important aspects of \emph{offline} social networking still remains largely unexplored. The analysis of such networks and the contained \emph{crowd intelligence} can potentially provide more direct answers to fundamental questions, \eg how do personal links get established, is it possible to correlate this with roles, how does personal communication evolve?

In this paper, we aim at providing first insights for answering such questions, also in order to leverage our analysis results for improving link prediction methods. We focus on real-world offline networks of \emph{human contacts}, that is, \emph{face-to-face} contacts between persons. In contrast to virtual networks, these contacts were acquired using a ubiquitous RFID-based system that allows us to collect individual face-to-face contacts. Thus, we can observe and analyze social interaction at a very detailed level, including the specific event sequences and durations.

For link prediction, we aim at predicting \emph{new} contacts based on network properties, as an adaptation of methods for online social networks. In addition, we extend the analysis in two important directions: First, we consider \emph{recurring} links: These are generated repeatedly in a network, \ie if a link between actors is formed multiple times. A prominent case of recurring links are face-to-face interactions. Second, we analyze influence factors and patterns for establishing such contacts, and also consider their specific \emph{durations} in a fine-grained dynamic analysis. Essentially, this also leads to a comprehensive analysis of the impact of \emph{weaker ties} for new and recurring contacts.

The context of our work is established by the social conference guidance system \conferator~\cite{conferator2010} implemented using the \textsc{Ubicon}~\cite{ABKSDHMMMS:14} platform (\url{http://www.ubicon.eu}). It provides ubiquitous access to conference information and allows conference participants to manage their contacts at the conference and to personalize their conference program. Using the system, conference participants can recall their individual contacts after the conference, \eg as virtual business cards. In addition, we apply link prediction for recommending \emph{interesting} contacts, \ie other participants. The system utilizes active RFID technology from the Sociopatterns project \url{http://www.sociopatterns.org} which allows us to analyze the collected contact (proximity) data between the participants as a proxy for their face-to-face contacts, as participatory sensing data of the whole conference. Using this data, we can derive special interaction networks, \ie contact networks, and apply those for link prediction.

For the analysis, we apply real-world data collected at three scientific conferences, \ie the LWA 2010 conference in Kassel, Germany, the Hypertext 2011 conference in Eindhoven, The Netherlands, and the LWA 2012 conference in Dortmund, Germany, using the \conferator system.
The results of the analysis indicate that weaker ties have a strong influence on the contact behavior and the prediction performance. We show, that there are clear influence patterns of the contact durations. Furthermore, we show that stronger links are better predictable than weaker links. Moreover, considering the contact durations in the ranking of the predicted contacts significantly improves the performance for the prediction of recurring links. This can be generalized for all three conferences.
Our contribution is summarized as follows:
\begin{compactenum}
  \item Concerning link prediction, we analyze the problem of predicting links in real-world \emph{human contact} networks, focusing on \emph{new} links.
  \item We adapt different state-of-the-art network proximity measures for the link prediction setting.
  \item We extend the basic link prediction problem, for predicting \emph{recurring} links.
  \item We compare neighborhood-based and path-based network proximity measures for the prediction of \emph{new} and \emph{recurring} links in networks of face-to-face proximity, using a threshold-based analysis for capturing contact dynamics; we show that stronger links are better predictable than weaker links. Moreover, we compare the performance of all considered measures to the \emph{current tie strength} predictor.
  \item Finally, we analyze the role of weak ties between actors, and show that they weaken the performance of the predictors.
\end{compactenum}

The remainder of this paper is structured as follows: Section~\ref{sec:related} discusses related work. After that, Section~\ref{sec:dataset} describes the RFID hardware setting and the collected datasets.
 Next, we present the used network proximity measures and link prediction techniques
 in Section~\ref{sec:measures}. We then discuss the results in Section~\ref{sec:analysis}.
Finally, Section~\ref{sec:conclusion} concludes with a summary and interesting options for future work.
This article is a significantly extended and revised version of \cite{SAS:12} and \cite{scholz2014predictability}.        

%% file: relatedwork.tex
\section{Related Work}\label{sec:related}
Before we focus on link prediction below, we first discuss related work concerning the analysis of human contact behavior, and the relevant connections.

\subsection{Analysis of Human Contacts}
Contacts patterns in social networks, and their underlying mechanisms are a classic topic of social network analysis. However, the analysis of offline social networks, focusing on human contacts, has been largely neglected. In this context,  Eagle et al. \cite{eagle2009} and
 Zhoe et al.~\cite{Hui2005} presented an analysis of proximity information collected by devices based on Bluetooth communication, similar to Xu et al.~\cite{XCWCZYWZ:11}, who also related this to online social networks.
 
 However, in all these experiments it was not possible to reliably detect face-to-face contacts. In contrast, in our experiments we use a new
generation of active RFID tags (proximity tags). The technical innovation of these tags is the possibility to detect the proximity
 of other tags, which allows us to recognize face-to-face contacts at a high detailed level including specific points in time and their durations.
 
One of the first experiments using proximity tags was conducted by Cattuto and colleagues in~\cite{ALANI09} at the ESWC 2009 conference.
 Here, the authors presented a novel application that combines online and offline data from the conference attendees. In \cite{Barrat10},
Cattuto and colleagues compared the attendees' contact patterns with their research seniority and their activity in social web platforms. They also extended their analysis to healthcare environments \cite{BarratCCIRTB10} and schools \cite{Stehl2011}. In \cite{MacekASS11}, we analyzed the dynamics of participants' contact pattern at conferences, and also the connection between academic jobs and roles at a conference.
In \cite{ADHMS:12,kibanov2013evolution, KASS14} the authors described the dynamics of community and contact structures. However, no analysis or application towards link prediction has been performed using the approaches discussed above. We discuss this important aspect in more detail below.
 
 \subsection{Link Prediction}
The prediction of new links between nodes in a social network is a challenging task.
A first comprehensive fundamental analysis was done by \cite{Kleinberg2003}.
Here, the authors defined the link prediction problem and studied link prediction approaches based on proximity measures of nodes in a co-authorship network of physics. \cite{Murata2007} analyzed weighted variants of different network proximity measures.
\cite{Wang2011} examined the impact of human mobility on link prediction.
\cite{LichtenwalterLC10} presented a new unsupervised (a restricted variant of rooted PageRank) and a new supervised method for the prediction of new links. \cite{BackstromL11} introduced a supervised method, based on supervised random walks, for the prediction of new links.

However, most of these approaches analyzed the predictability of new links in online social networks like Facebook or DBLP. The prediction of links in offline social networks has been largely neglected. For reliably detecting face-to-face proximity, a new generation of active RFID tags has been developed by the SocioPatterns collaboration, cf.~\cite{CIROHIGHRES08}, which we also applied for the data analyzed in this work. In~\cite{SAS:12}, we presented a first analysis concerning the predictability of new and recurring links in real world face-to-face contact networks. In~\cite{scholz2013insights}, we further showed that the predictability of new links can be further improved by data from online networks, proposing a new unsupervised link prediction method that combines the information of different networks. In~\cite{scholz2013people}, we analyzed triadic closure in face-to-face contact networks. In~\cite{Tsugawa2013}, the authors also analyzed the quality of unsupervised methods in the context of link prediction in face-to-face proximity networks; they compared the predictability of links in face-to-face contact networks and other types of social networks, supporting our earlier work presented in~\cite{SAS:12}. 

%% file: rfid-dataset.tex
\section{Face-to-Face Contact Data}\label{sec:dataset}


In the following section, we first describe the applied active RFID technology for collecting the data. Next, we present the three datasets collected at the LWA 2010,\footnote{\url{http://www.kde.cs.uni-kassel.de/conf/lwa10/}} the Hypertext 2011\footnote{\url{http://www.ht2011.org/}} and the LWA 2012\footnote{\url{http://lwa2012.cs.tu-dortmund.de/}} conferences, and provide initial characteristic statistics. After that, we define our problem setting for link prediction, and describe, how we model the underlying networks.


\subsection{RFID Datasets}

For our experiments we asked each conference participant to wear an active RFID tag. These so called proximity tags
 are developed by the SocioPatterns project. One decisive factor of these tags is the possibility to detect other proximity tags
within a range of up to 1.5 meters which allows us to identify and analyze
 human face-to-face contacts. Each RFID tag sends signals to RFID readers that are placed at fixed positions in the conference area.
The RFID readers forward these signals to a central server, where all signals are stored into a database. Each signal contains the ID of the
 transmitting tag and the IDs of all RFID tags in its proximity.
For more information about the proximity tags we refer to~\cite{CIROHIGHRES08} and the OpenBeacon website (\url{http://www.openbeacon.org}).

\begin{figure}
    \centering
      \includegraphics[width=0.5\columnwidth]{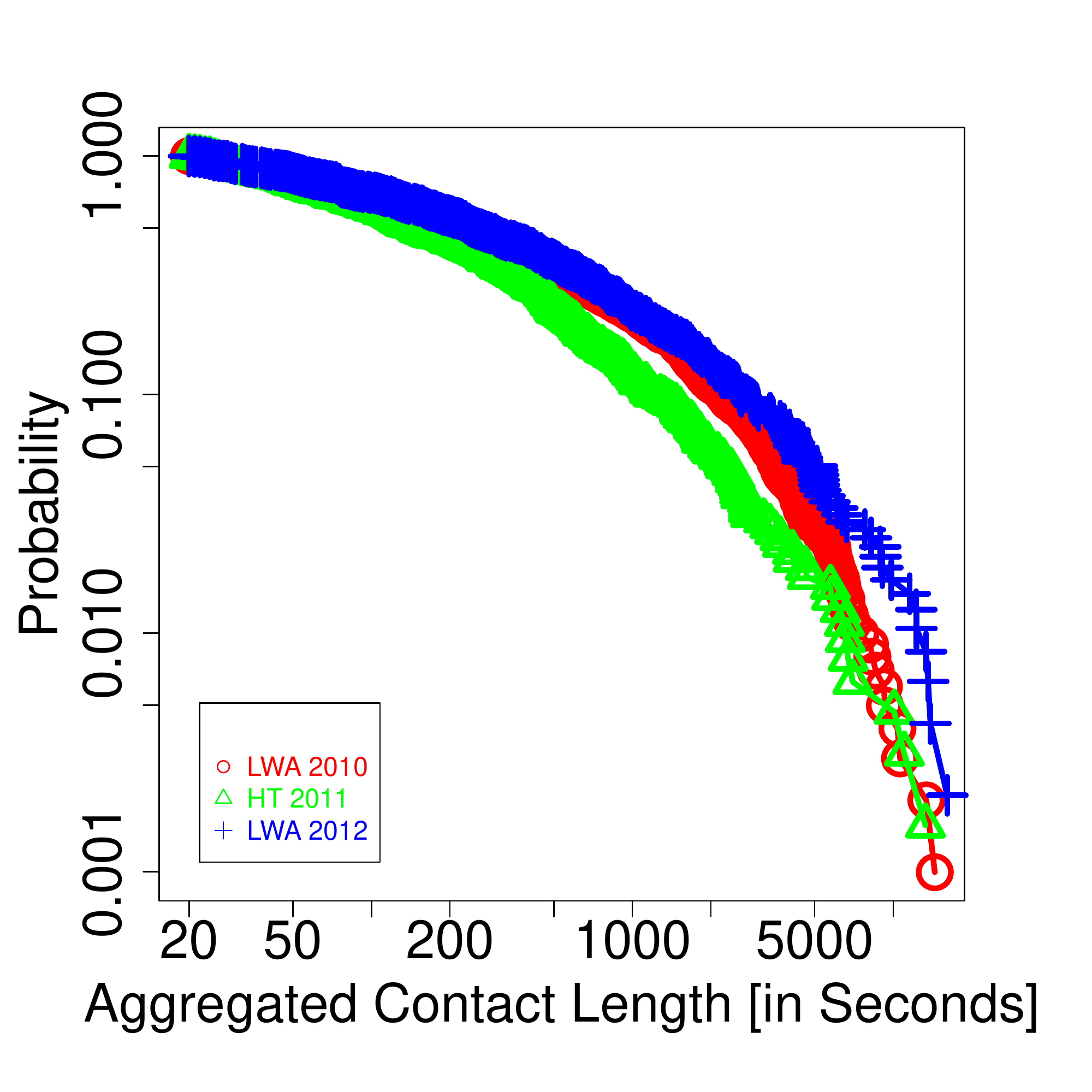}
  \caption{Aggregated contact length distribution of all aggregated face-to-face contacts of the LWA 2010,  the HT 2011 and the LWA 2012 conferences, respectively: The $x$-axis displays the minimum length of an aggregated contact in seconds, the $y$-axis the probability for an aggregated face-to-face contact having at least this contact length, respectively. Both axes are scaled logarithmically.}
  \label{fig:contactlengthdistribution}
\end{figure}

Table \ref{tab:datasetstatistics} gives a detailed description of the collected datasets. For the LWA 2010, Hypertext (HT) 2011, and LWA 2012 conferences, we used the first day of the conference as training data. Hence, we aim to predict new and recurring contacts of day two and three. In Figure~\ref{fig:contactlengthdistribution}, we observe the typical distribution of all face-to-face contacts for all three conferences. 
Confirming previous findings, e.g. in \cite{DBLP:journals/corr/abs-1006-1260,ADHMS:12,MacekASS11}, 
most of the aggregated contacts take less than two minutes and the aggregated contact durations of both conferences show a long-tailed distribution. In addition, Figure~\ref{fig:contactlengthdistribution} shows, that the number of long aggregated face-to-face contacts at LWA 2012 was significantly higher than at HT 2011 and LWA 2010. The diameter and average path length of $G$ is similar to the results presented in \cite{DBLP:journals/corr/abs-1006-1260,ADHMS:12}.

\begin{table}[htb]
\footnotesize
\centering
 \renewcommand{\arraystretch}{0.9}
   \caption{General statistics for the collected datasets. Here $d$ is the diameter, \textit{APL} the average path length, \textit{ACL} the average contact length and \textit{LCN} the largest clique number.
   \label{tab:datasetstatistics}}
    {
    \begin{tabular}{|l|c|c|c|c|}
    \hline
    & \textbf{LWA 2010} & \textbf{HT 2011} & \textbf{LWA 2012}\\
    \hline\hline
    \textit{\#days} & $3$ & $3$ & $3$\\
    \hline
    \textit{$|V|$} & $77$ & $62$ & $42$\\
    \hline
    \textit{$|E|$} & $1004$ & $640$ & $478$\\
    \hline
    \textit{Avg.Deg.($G$)} &  $26.06$ & $20.53$ & $22.76$\\
    \hline
    \textit{APL ($G$)} &  $1.7$ & $1.7$ & $1.45$\\
    \hline
    \textit{ACL ($G$)} &  $797.50$ & $529.07$ & $1023.17$\\
    \hline
    \textit{LCN ($G$)} &  $4$ & $5$ & $5$\\
    \hline
    \textit{d ($G$)} &  $3$ & $3$ & $3$ \\
    \hline
    \textit{$|V_{\mathrm{core}}|$} & $57$ & $49$ & $32$\\
    \hline
    \textit{$|E^{\le t}|$} & $426$ & $481$ & $263$\\
    \hline
    \textit{$E^{>t}_\mathrm{core}\setminus E^{\leq t}$} &  $394$ & $132$ & $134$\\
    \hline
    \textit{$E^{>t}_\mathrm{core} \cap E^{\le t}$} &  $242$ & $134$ & $115$\\
    \hline
    \textit{Avg.Deg.($G(\le t)$)} &  $27.04$ & $32.1$ & $27.68$\\
    \hline
    \textit{APL ($G(\le t)$)} &  $1.9$ & $1.84$ & $1.65$ \\
    \hline
    \textit{LCN ($G(\le t)$)} &  $4$ & $4$ & $4$\\
    \hline
    \textit{d ($G(\le t)$)} &  $4$ & $4$ & $3$\\
    \hline
    \textit{Avg.Deg.($G(> t)$)} &  $44.6$ & $21.7$ & $31.12$\\
    \hline
    \textit{APL ($G(> t)$)} &  $1.64$ & $1.99$ & $1.53$\\
    \hline
    \textit{LCN ($G(> t)$)} &  $5$ & $4$ & $4$\\
    \hline
    \textit{d ($G(> t)$)} &  $3$ & $4$ & $3$\\
    \hline
    \end{tabular}
    }
\end{table}

\subsection{Problem Statement}

Let $t$ be a point in time during the conference. For the prediction task, we consider all face-to-face contacts starting before $t$ as training data and face-to-face contacts starting later as test data.

\vspace*{.1em}
The \textbf{training data} is thus the undirected graph $G^{\leq t}=(V^{\leq t},E^{\leq t})$, where $V^{\leq t}$ is the set of all participants who had at least one face-to-face contact with some other participant before $t$; two participants $u,v\in V^{\leq t}$ are connected by an edge  $(u,v) \in E^{\leq t}$, if they had at least one face-to-face contact before $t$. The weight $w_{\le t}(u,v)$ is the sum of the durations of all their face-to-face contacts before $t$.
%
The \textbf{test data} is the undirected graph $G^{>t}=(V_\mathrm{core}, E^{>t}_\mathrm{core})$, where $V_\mathrm{core}$ is the set of participants who had at least one contact during the training interval and at least one contact during the test interval: Two participants $u,v \in V_\mathrm{core}$ are connected by an edge  $(u,v) \in E^{>t}_\mathrm{core}$ if $u$ and $v$ had at least one face-to-face contact after $t$. The weight $w_{>t}(u,v)$ sums up the durations of all their face-to-face contacts after $t$.

\vspace*{.9em}
\noindent\textbf{Prediction Tasks.}
In this paper, we consider the following link prediction tasks: 
 \begin{enumerate}
 \item New links only (as in \cite{Kleinberg2003}), \ie all links in $E^{>t}_\mathrm{core}\setminus E^{\leq t}$.
 \item Recurring links, \ie all links in $E^{>t}_\mathrm{core} \cap E^{\le t}$.
 \end{enumerate}

Note that --- following the approach in \cite{Kleinberg2003} --- the training set $G^{\leq t}$ contains all vertices of $G$, while the test set $G^{>t}$ contains only vertices $v \in V_\mathrm{core}$, \ie those that are present in the core.


%% file: measures.tex
\section{Network Proximity Measures}\label{sec:measures}
In this section, we discuss neighborhood-based and path-based measures used in our analysis for the prediction tasks.
Focussing on unsupervised methods, most of the predictor scores are based on either nodes' neighborhoods or path information. All of these proximity measures are based on the assumption that two nodes have a higher probability to become connected, if these two nodes are close in the graph.

\subsection{Neighborhood-based Network Proximity Measures}
In Table \ref{tab:networkprosimitymeasures}, we provide a detailed overview of the used unweighted and weighted proximity measures.
 The measure \emph{Common Neighbors} is based on the assumption that it is more likely that two nodes are connected if these two nodes have many neighbors in common. \emph{Adamic Adar} and \emph{Resource Allocation} are similar to \emph{Common Neighbors}, but here the \emph{Common Neighbors} are weighted with respect to their degree. Considering \emph{Jaccard's Coefficient} it is more likely that two nodes are connected, if these two nodes share a high fraction of their respective neighborhood.  
 \emph{Preferential Attachment} is based on the assumption, that the probability  \cite{barabasi2002} of a new node being connected to node $x$ is proportional to the degree of $x$.
We define the neighborhood for a node $x$, \ie the set of neighbors $N(x)$, as
\begin{equation*}
 N(x) = \{y | y \in V, (x,y) \in E\}
\end{equation*}


\begin{table*}[thb]
\centering
 \renewcommand{\arraystretch}{0.95}
 \renewcommand\tabcolsep{3pt}
       \caption{Overview of network proximity measures based on the nodes' neighborhood.
       \label{tab:networkprosimitymeasures}}{
    \begin{tabular}{|l|c|c|}
    \hline
    \textbf{Measure} & \textbf{Unweighted} & \textbf{Weighted}\\
    \hline\hline
    \textit{Common Neighbors} & $\mathit{CN}(x,y) = |N(x) \cap N(y)|$ & $\mathit{WCN}(x,y) =$ \(\sum\limits_{z \in N(x) \cap N(y)} w(x,z) + w(y,z)\)\\
    \hline
    \textit{Adamic-Adar } & $\mathit{AA}(x,y) = \sum\limits_{z \in N(x) \cap N(y)} \frac{1}{\log{|N(z)|}}$ & $\mathit{WAA}(x,y) = \sum \limits_{z \in N(x) \cap N(y)} \frac{w(z,x) + w(z,y)}{\log{(\sum\limits_{z^{'} \in N(z)} w(z,z^{'}))}}$\\
    \hline
    \textit{Jaccard's Coefficient} & $\mathit{JC}(x,y) = \frac{|N(x) \cap N(y)|}{|N(x) \cup N(y)|}$  & $\mathit{WJC}(x,y) = \frac{\sum \limits_{z \in N(x) \cap N(y)} w(x,z) + w(y,z)}{\sum \limits_{x^{'} \in N(x)} w(x,x^{'}) + \sum \limits_{y^{'} \in N(y)} w(y,y^{'})}$\\
    \hline
    \textit{Resource Allocation } & $\mathit{RA}(x,y) = \sum\limits_{z \in N(x) \cap N(y)} \frac{1}{|N(z)|}$ & $\mathit{WRA}(x,y) = \sum \limits_{z \in N(x) \cap N(y)} \frac{w(z,x) + w(z,y)}{\sum\limits_{z^{'} \in N(z)} w(z,z^{'})}$\\
    \hline
    \textit{Pref. Attachment} & $\mathit{PA}(x,y) =  |N(x)|\cdot |N(y)| $ & $\mathit{WPA}(x,y) = \sum \limits_{x^{'} \in N(x)} w(x,x^{'}) \cdot \sum \limits_{y^{'} \in N(y)} w(y,y^{'})$\\
    \hline
    \end{tabular}
    }
\end{table*}

\subsection{Path-based Network Proximity Measures}
 The \emph{rooted PageRank} \cite{Kleinberg2003} predictor is an adaption of the PageRank algorithm \cite{BrinP98} for the link prediction task. The \textit{rooted PageRank} algorithm computes the stationary probability distribution of participant $y$ under the following random walk~\cite{Kleinberg2003}:
\begin{itemize}
\item With probability $\alpha$, jump to $x$.
\item With probability $1-\alpha$, jump to a random neighbor of the current node.
\end{itemize}

The \emph{Katz}~\cite{Katz53} predictor is defined as 

 \begin{equation*}
  \mathit{Katz}(x,y) = \sum_{l=1}^{\infty} \beta^l \cdot |\text{path}_{x,y}^l|,             
 \end{equation*}
  where $\text{path}_{x,y}^l$ is, for $x,y \in V$, the set of paths from $x$ to $y$ with length $l$.
 We note that $\beta \in [0,1]$ is a damping factor that weights short paths higher/lower in the summation. 

%% file: analysis.tex
\section{Analysis}\label{sec:analysis}

In this section, we study the link prediction problem on human contact networks. As already done in literature~\cite{Wang2011,Kleinberg2003,Murata2007}, we analyze the predictability of several network proximity measures (see Table \ref{tab:networkprosimitymeasures}).
In contrast to previous work, we also extend our studies to the prediction of recurring links, and analyze the influence of weak ties. Furthermore, we study the predictability of stronger ties concerning the new and recurring link prediction problem. We start with some statistics about the contact behavior of participants at the three conferences.

\subsection{Human Communication Statistics and Basic Analysis}
\label{sec:analysis:basic}
 
\begin{figure}[!h]
      \begin{center}
      \includegraphics[width=.62\columnwidth]{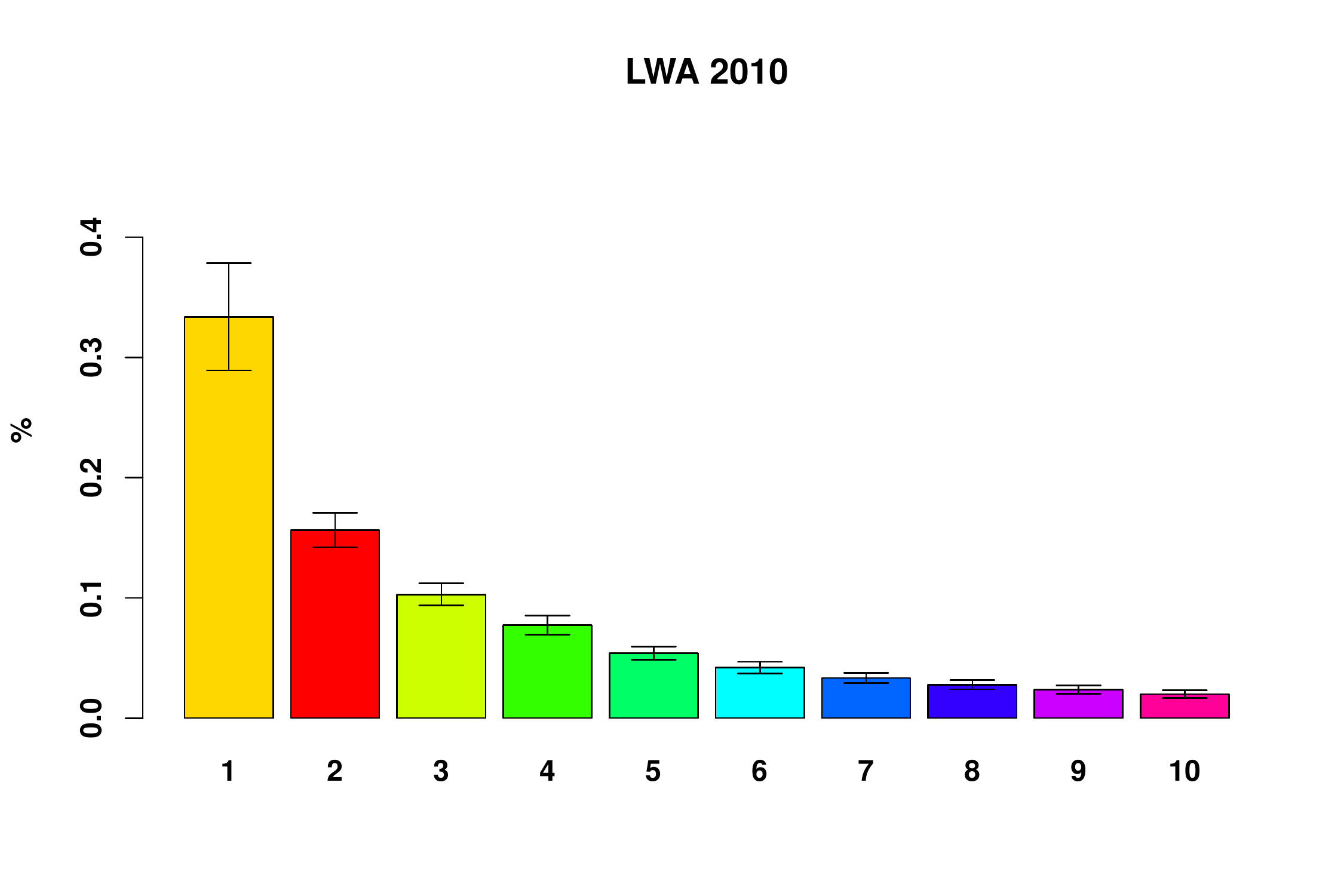}
      \includegraphics[width=.62\columnwidth]{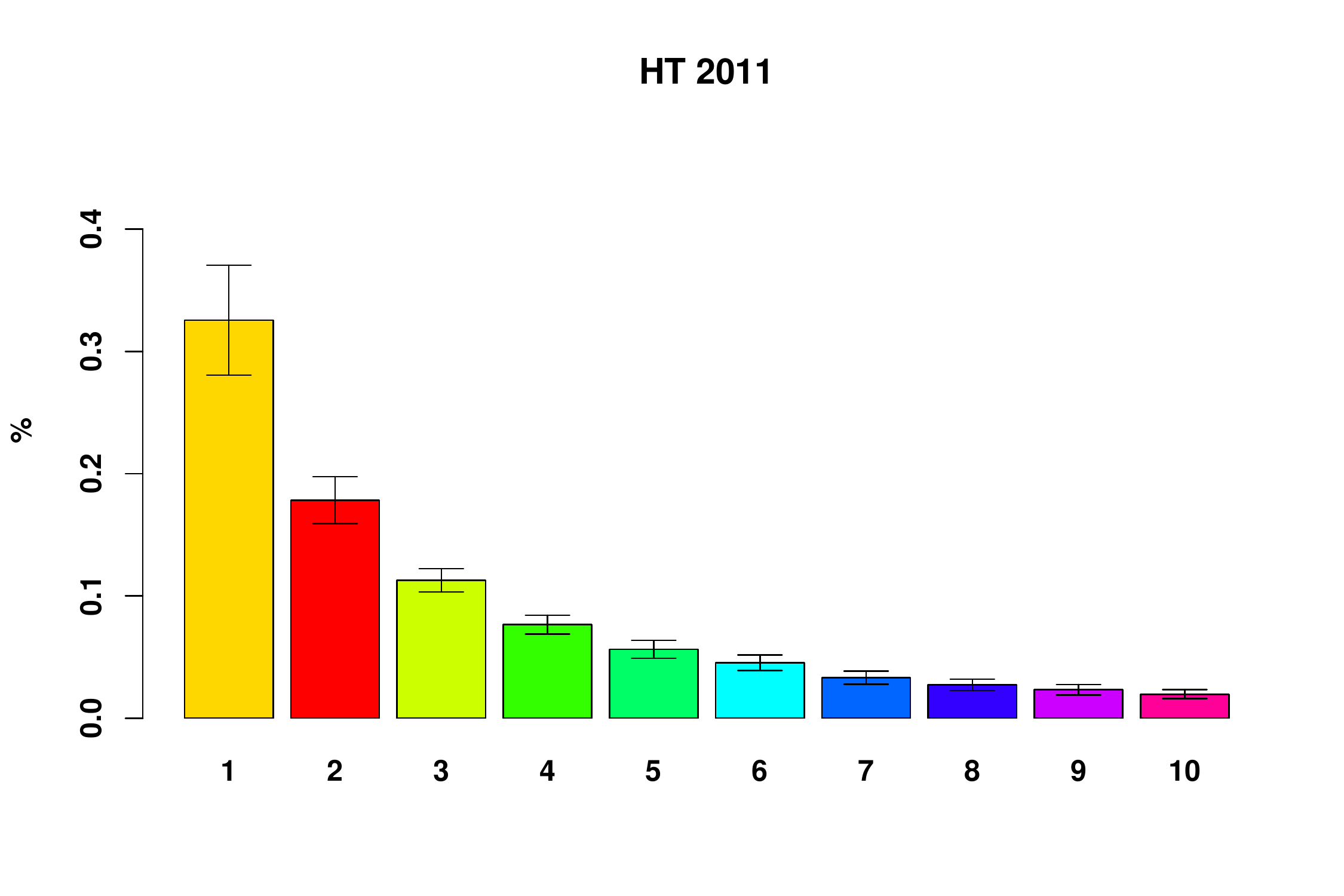}
    \includegraphics[width=.62\columnwidth]{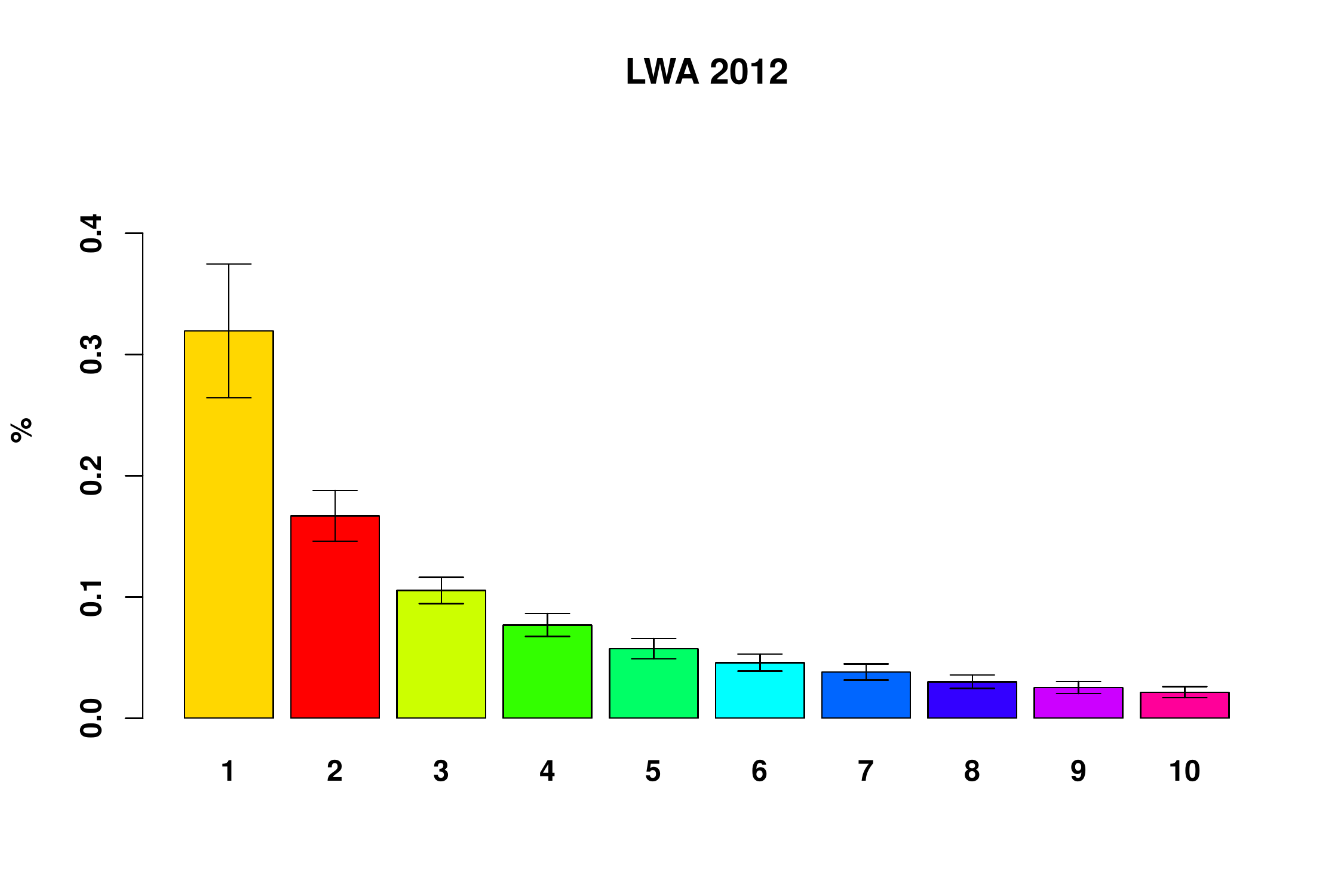}
      \end{center}
  \caption{Average fraction of contact duration to each participant's longest, second longest, $\ldots$, tenth longest face-to-face contact, for each conference. The $x$-axis represents the $i-$th longest  contact; the $y$-axis shows the fraction of contact duration to the $i-th$ longest contact. Here, for example,
 the left bar (labeled with $1$) in the HT 2011 bar-plot means, that in average the fraction of a participant's longest face-to-face contact partner is approximately $33$ percent of his total contact duration. The error bars indicate the 95 percent confidence interval.}
  \label{fig:LongestContactsDistribution}
\end{figure}

Knowledge about human communication behavior is important to improve the prediction of future links. We therefore present some new insights into
 the communication behavior of participants during a conference. In Figure~\ref{fig:LongestContactsDistribution}, we analyze the average 
 contact length distribution with the longest, second longest, $\ldots$, tenth longest contact. On average, each participant's longest face-to-face contact 
is at least one third of his 
 total face-to-face contact time. This fraction decreases rapidly (from 33 percent for the longest contact), when we consider the fraction of the second longest contact. Here, the fraction of the contact length
 compared to the overall contact length is approximately 17 to 19 percent.
 Interestingly, all bar-plots look quite similar at all conference datasets. This might indicate, that this is the typical behavior in a conference setting.
 
 \begin{figure}[!htb]
\begin{center}
      \includegraphics[width=.497\columnwidth]{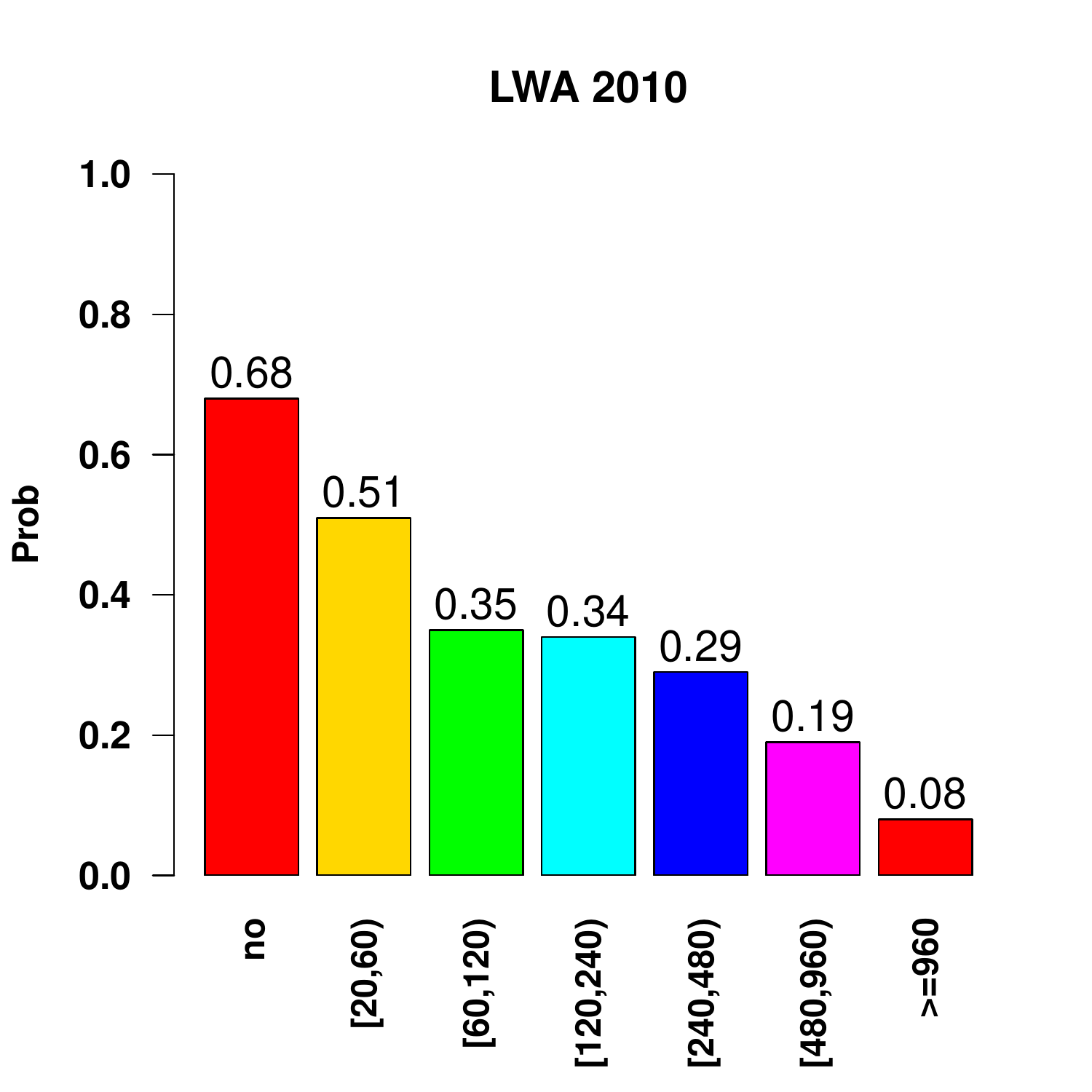}
      \includegraphics[width=.497\columnwidth]{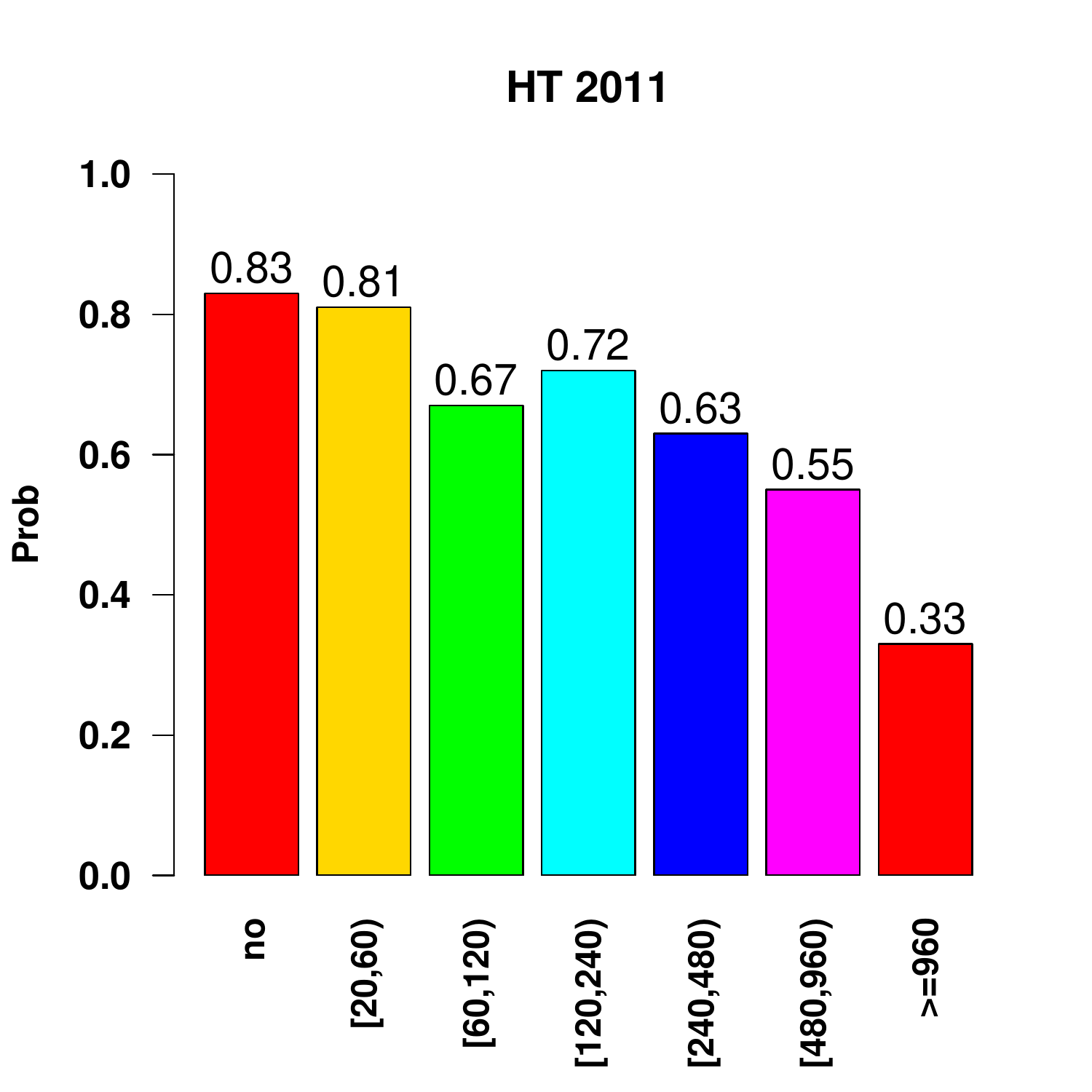}
      \includegraphics[width=.497\columnwidth]{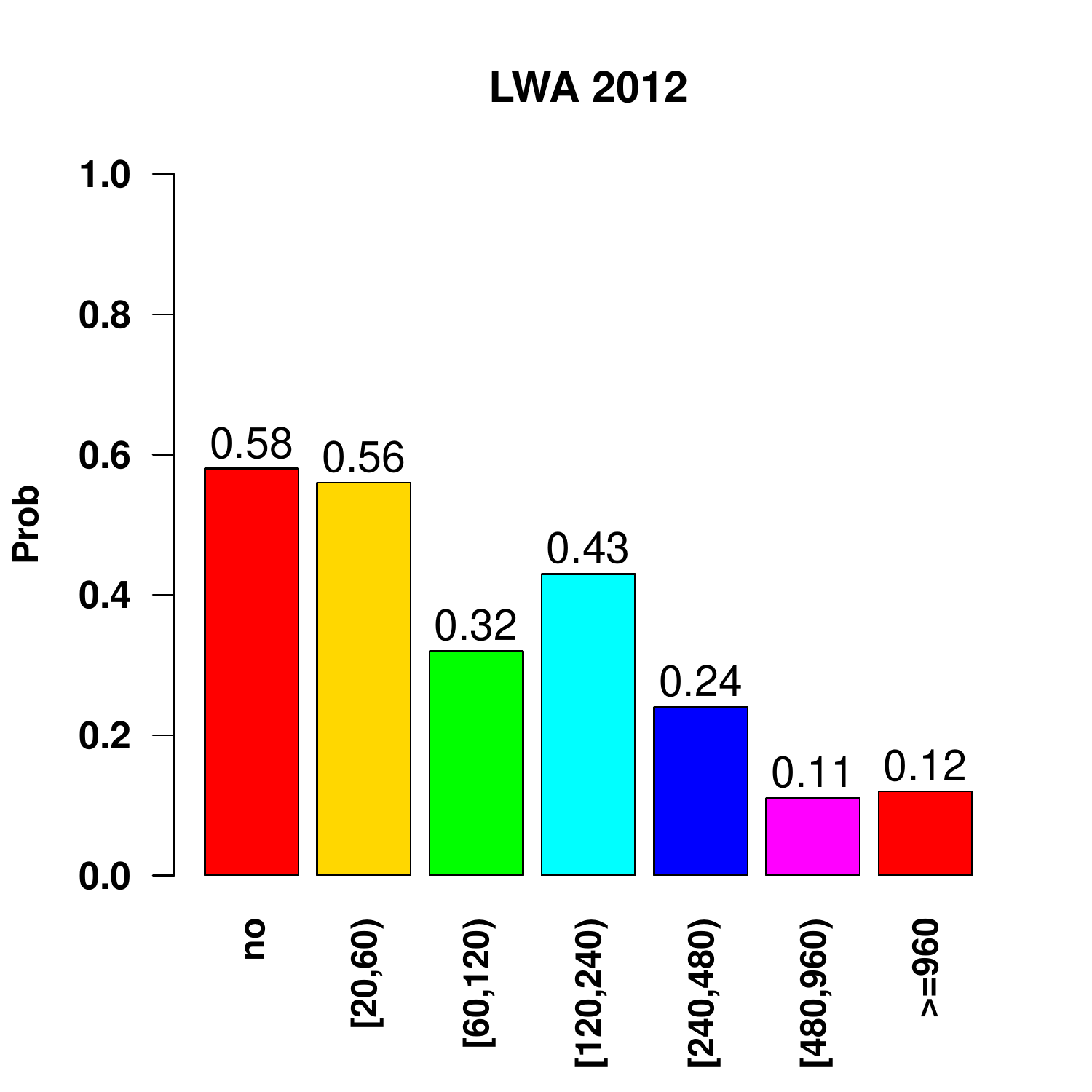}
  \caption{Impact of contact duration between two participants on the first day on the contact between these two participants for the remaining days (day two and three), for all conferences.
 The left bar labeled with 'no' in the HT 2011 bar-plot means, for example, that two participants who had no contact at the first day of the conference
 had no contact until the end of the conference in $83.5 \%$. The bar labeled with $[20,60)$ in the HT 2011 bar-plot means that two
 participants who had a contact with a duration between $20$ and $60$ seconds on the first day had no contact on the second and third day 
in $80.6\%$ of all cases.
In Table \ref{tab:nocontacts} we present the detailed numbers for these figures. The column $\sum$ Contacts 
 represents the number of contacts for the specified type of contact. Here for example (for HT 2011) the row 'no' means, that there are
 1230 pairs of participants who had no contact at the first day. 836 of these pairs had no contact at the second and third day, either. The row $[20,60]$ means
 that there are 110 participants, who had a contact with contact duration between 20 and 60 seconds. 56 of these had no further contact on the second and third day.} 
  \label{fig:noContacts}
\end{center}
\end{figure}

\begin{table*}[!htb]%
\footnotesize 
\centering
\renewcommand{\tabcolsep}{2.4pt}
\renewcommand{\arraystretch}{0.90}
  \caption{Overview about the number of participants for different time intervals used in Figure~\ref{fig:noContacts}.
  \label{tab:nocontacts}}{
  \begin{tabular}{|l|c|c|c|c|c|c|}
      \hline
      & \multicolumn{2}{l||}{\textbf{LWA 2010}} & \multicolumn{2}{l|}{\textbf{HT 2011}} & \multicolumn{2}{l|}{\textbf{LWA 2012}}\\\cline{1-7}
      & \scriptsize $\sum$ Contacts & \scriptsize $\#no$ Contacts & \scriptsize $\sum$ Contacts & \scriptsize $\#no$ Contacts &\scriptsize $\sum$ Contacts & \scriptsize $\#no$ Contacts\\
        \hline\hline
      no & 1230 & 836 & 798 & 666 & 320 & 186\\\hline
      $[20,60)$ & 110 & 56 & 98 & 79 & 45 & 25\\\hline
      $[60,120)$ & 63 & 22 & 64 & 43 & 34 & 11\\\hline
      $[120,240)$ & 56 & 19 & 60 & 43 & 30 & 13\\\hline
      $[240,480)$ & 58 & 17 & 62 & 39 & 33 & 8\\\hline
      $[480,960)$ & 31 & 6 & 40 & 22 & 18 & 2\\\hline
      $\ge 960$ & 48 & 4 & 54 & 18 & 16 & 2\\\hline
 \end{tabular}
 }
\end{table*}

\begin{figure}[!htb]
      \begin{center}
      \includegraphics[width=.495\columnwidth]{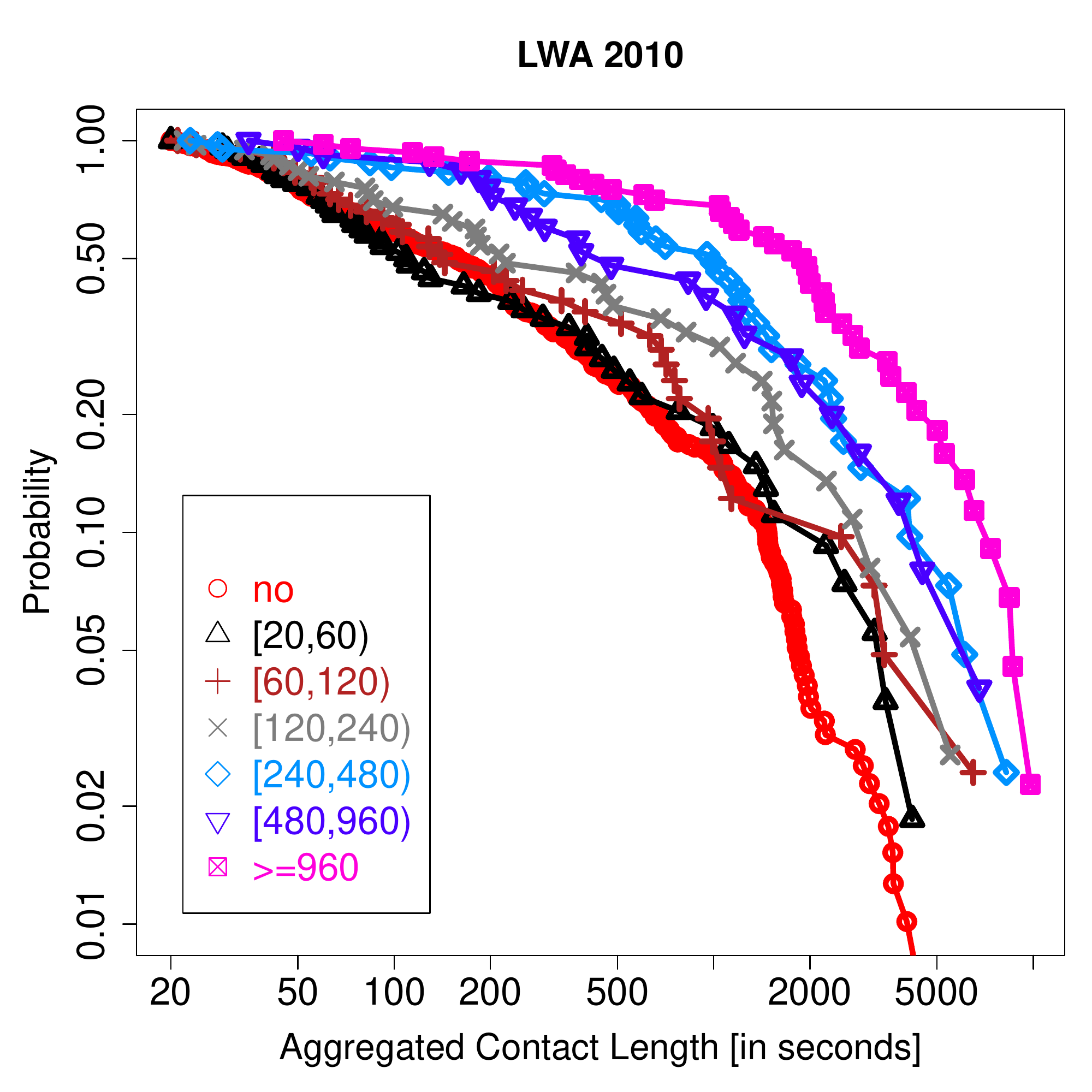}     
      \includegraphics[width=.495\columnwidth]{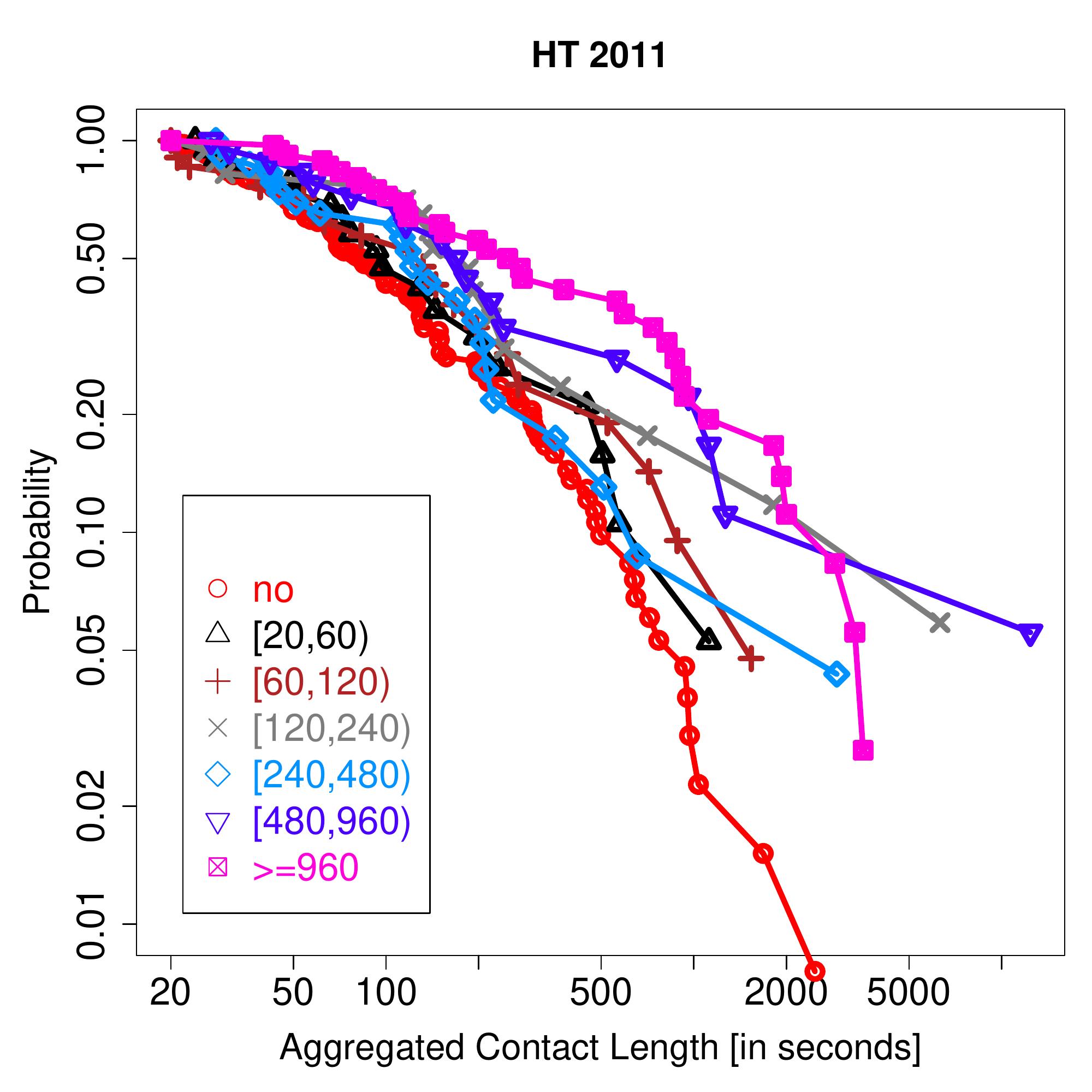}
     
       \includegraphics[width=.495\columnwidth]{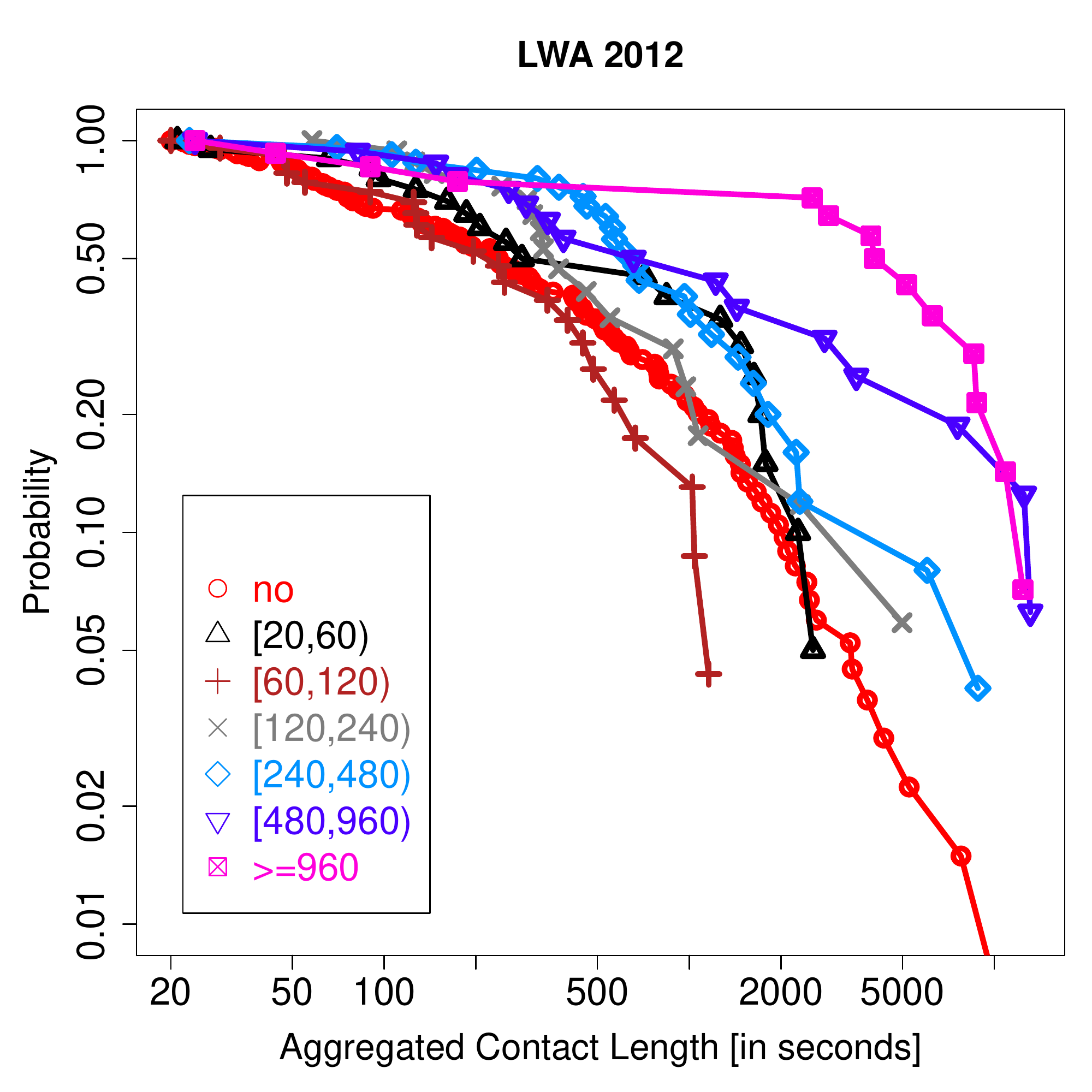}
      \end{center}

  \caption{Impact of the contact duration between two participants on the first day, and the 
 contact length of a recurring contact at the second and third day, for all conferences. The red line labeled with 'no' (circle symbol) in the LWA 2010 plot, for example, shows the distribution of all contacts between participants at day two and three, which had no
 contact at the first day. The line labeled with $[60,120)$ (cross symbol) shows the distribution of all contacts between participants at day two and three, which had a contact
 with contact duration between 60 and 120 seconds at the first day of the conference.} 
  \label{fig:impactContactLength}
\end{figure}

 Will participants who had a contact at the first day of the conference recur this contact  again on the second or third day? For answering this question and for understanding its mechanisms it is important to consider the contact length from the first day of  the conference. In Figure~\ref{fig:noContacts}, we observe the clear trend, that a contact is more likely to be renewed the longer the contact on the first day. In Figure~\ref{fig:impactContactLength}, we plot the distribution of all contacts for the second and third day, depending on the contact length of the first day. 
We observe, that a longer contact is more likely, the longer the contact on the first day. An interesting further question is then to find typical features to predict renewed contacts and their length.

\subsection{Role-based Influence Factors}

In the following, we analyze the impact of a number of (external) role-based factors for the link-prediction problem, relating to properties of the people collaborating in the contact network. Specifically, we focus on the prediction of new contacts and recurring links.

We use pattern mining for identifying \emph{characteristic patterns}~\cite{ALKH:09b,AL:12a} describing subgroups with a high share of \emph{new contacts}. The applied technique is subgroup discovery, e.g.,~\cite{AP:06,APB:05a,Wrobel:97}: Basically, we aim at discovering subgroups of participants described by combinations of factors, e.g., \emph{session chair AND strong affiliation} that show a high share of a certain target property, an increased mean of new contacts compared to the default share.
Intuitively, we identify conjunctions of attribute values describing subsets of a dataset that maximize a given property, e.g., an increased mean of an attribute in the subset compared to the whole dataset. In the patterns described below, this \emph{target attribute} is given by the \emph{mean contact count} of new contacts.

\begin{table}[h]%
\footnotesize 
\centering
\renewcommand{\arraystretch}{0.90}
  \caption{Partitions of the set of participants into subgroups according to academic status and affiliation with HT 2011.
  \label{tab:communitystatistics}}{
    \begin{minipage}[t]{0.45\linewidth}
    \begin{tabular}{|l|c|}
    \hline
    \multicolumn{2}{|l|}{\textbf{Academic Status}}\\\hline
    Professor & 14\\\hline
    PhD-candidate & 34 \\\hline
    PhD  & 20\\\hline
    Other & 7\\\hline
    \end{tabular}
    \end{minipage}
    \begin{minipage}[t]{0.25\linewidth}
    \begin{tabular}{|l|c|}
    \hline
    \multicolumn{2}{|l|}{\textbf{Affiliation with HT}}\\\hline
    high & 12\\\hline
    medium & 17 \\\hline
    low  & 46\\\hline
    \end{tabular}
    \end{minipage}
}
\end{table}
\begin{table}[h]%
\footnotesize 
\centering
\renewcommand{\arraystretch}{0.90}
    \caption{Exemplary top 5 role influence patterns for the Hypertext 2011 conference measuring the increase in new contacts.
    \label{tab:subgroups:new:ht2011}}{
\begin{tabular}{|l|c|c|c|r|}
\hline
 \# & Lift & Mean & Size & Description\\
\hline
1 & 1.58 & 8.50 & 6 & session chair AND strong affiliation\\
2 & 1.55 & 8.36 & 11 & professor\\
3 & 1.35 & 7.25 & 8 & session chair\\
4 & 1.31 & 7.08 & 12 & strong affiliation\\
5 & 1.08 & 5.81 & 16 & presenter\\ 
\hline
\end{tabular}
}
\end{table}
\begin{table}[h]%
\footnotesize 
\centering
\renewcommand{\arraystretch}{0.90}
    \caption{Exemplary role influence patterns for the Hypertext 2011 conference measuring the mean of recurring contacts.
    \label{tab:subgroups:recurring:ht2011}}{    
\begin{tabular}{|l|c|c|c|r|}
\hline
 \# & Lift & Mean & Size & Description\\
\hline
\hline
1 & 2.10 & 5944.17 & 6 & PhD AND low affiliation\\
2 & 1.52 & 4297.15 & 26 & low affiliation\\
3 & 1.09 & 3089.00 & 6 & session chair AND professor\\
4 & 1.08 & 3038.67 & 21 & PhD candidate\\
5 & 1.06 & 3003.93 & 14 & PhD\\
6 & 0.87 & 2461.25 & 8 & session chair\\
7 & 0.82 & 2326.18 & 11 & professor\\
\hline
\end{tabular}
}
\end{table}

%


We focused on different subgroup structures, \ie partitionings, 
 induced by academic status, affiliation with the Hypertext conference series, 
and affiliation with one of the four conference tracks. 
In Table~\ref{tab:communitystatistics}, we present some
 statistics about the different subgroups.
 We classify participants as highly affiliated with the Hypertext conference series if they presented a paper
 more than three times at Hypertext conferences in different years.  The affiliation of a participant is low when he or she has never
presented a paper or presented a paper at Hypertext 2011 for the first time.  All other participants are classified with a medium affiliation.
The tables show the lift of the pattern assessing the ratio of the mean of new contacts covered by the pattern and the fraction of the whole dataset, the size of the pattern extension (number of described participants), and the description itself. The first line in Table~\ref{tab:subgroups:new:ht2011}, for example, shows that being a session chair with a strong affiliation to HT 2011 increases the mean number of new contacts by $58\%$.

Below, we exemplarily show interesting patterns with respect to the Hypertext 2011 conference. We collected conference and participants roles and analyzed their correlation with the emergence of new contacts. As shown in Table~\ref{tab:subgroups:new:ht2011}, as expected we observe an influence of being a session chair at the conference; this is even increased for participants with stronger affiliation to the conference, i.e., if participants are more experienced and also have more publications at previous conferences.
As expected, we observe that \emph{presenters} encounter a lot of new contacts. Also, the academic status of \emph{Professor} increases the contact count, as also confirmed by the LWA 2010 data.

%
%
%
%
%

In addition to new contacts, we also analyzed recurring contacts and their contact durations. Table~\ref{tab:subgroups:recurring:ht2011} shows exemplary patterns for the Hypertext 2011 conference. While we observe, that people with a low affiliation, \ie participants that are new to the conference are still very active after the first day, an interesting finding for Hypertext is, that being a session chair and being a professor increases the mean duration of contacts by $10\%$ while the single factors alone inhibit the duration ($-13\%$ and $-18\%$, respectively). For the LWA 2010, for example, we found a slightly different pattern; the organizers were still very active (increase by $34\%$), but the professors scored as expected (increase by $17\%$).

\subsection{Evaluation Method}\label{sec:linkprediction:evaluation}
For the evaluation of link prediction measures, often the precision of the top $n$ predicted links is used \cite{Kleinberg2003}, where $n$ is the number of positive links 
(i.e. the number of new or renewed links on day 2 and 3). In this work, we measure the accuracy by the area under a receiver operating characteristic (AUC) \cite{Hanley1982}. 
 In short, receiver operating characteristic (ROC) graphs plot the true positive rate on the $y$-axis and the false positive rate on the $x$-axis, concerning the set of predictions (ranking).
 The advantage of AUC is that it considers the whole ranking. For the prediction of new or recurring links each network proximity measure (predictor) outputs a ranked list in decreasing order of confidence. 

\subsection{Prediction of New Links}\label{sec:linkprediction:newLinks}
\begin{figure}[htb]
\begin{center}
      \includegraphics[width=0.4\columnwidth]{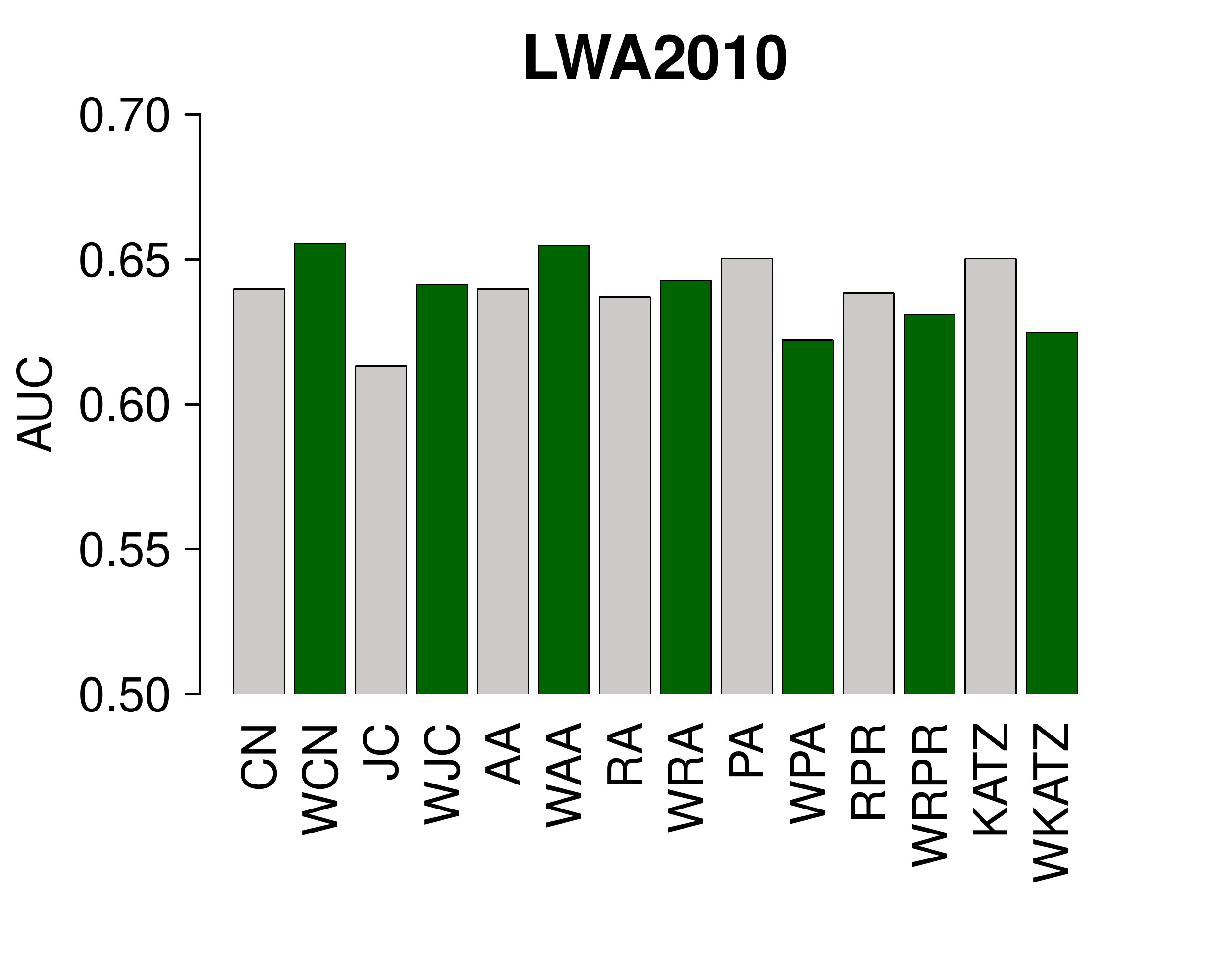}
      \includegraphics[width=0.4\columnwidth]{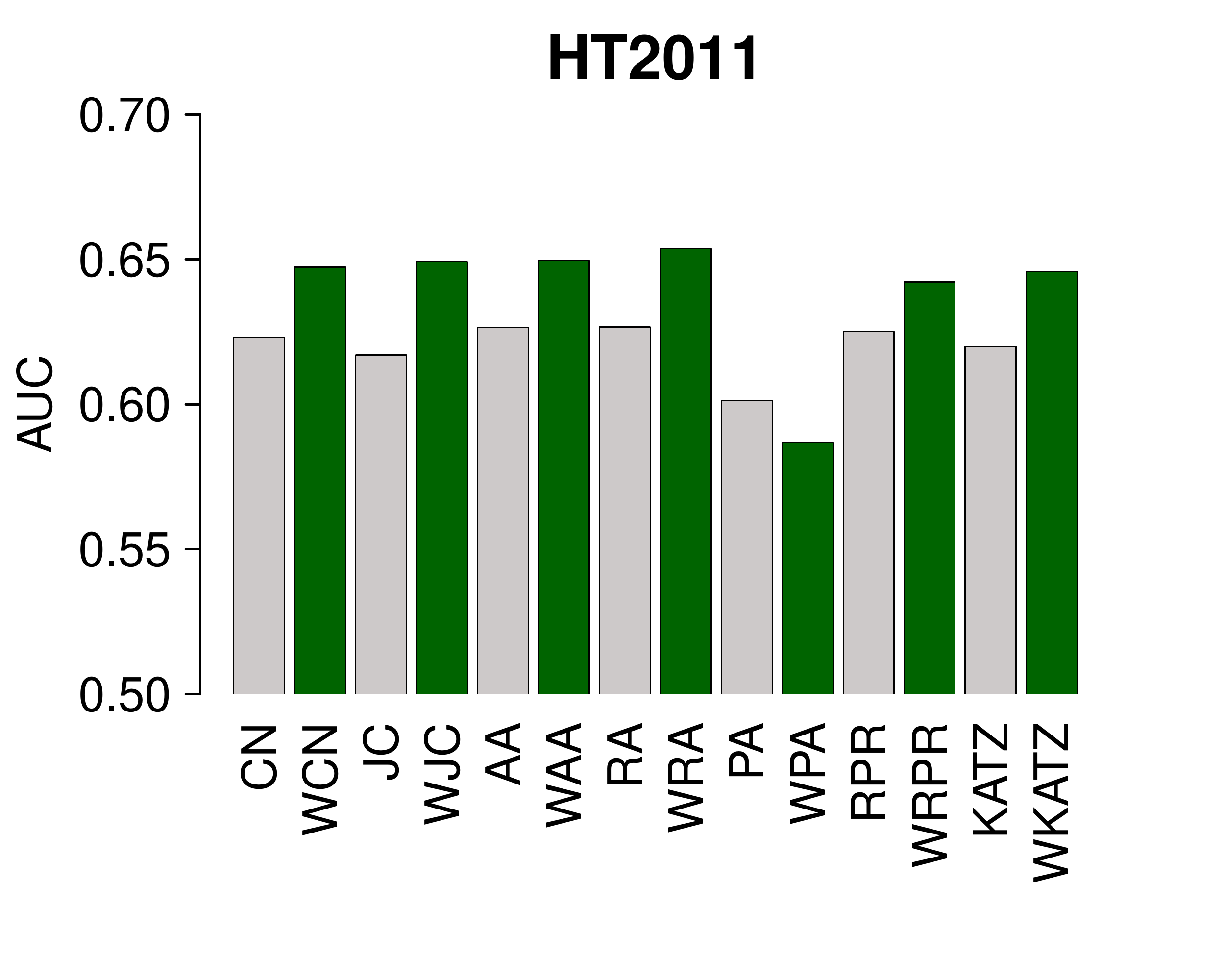}
      \includegraphics[width=0.4\columnwidth]{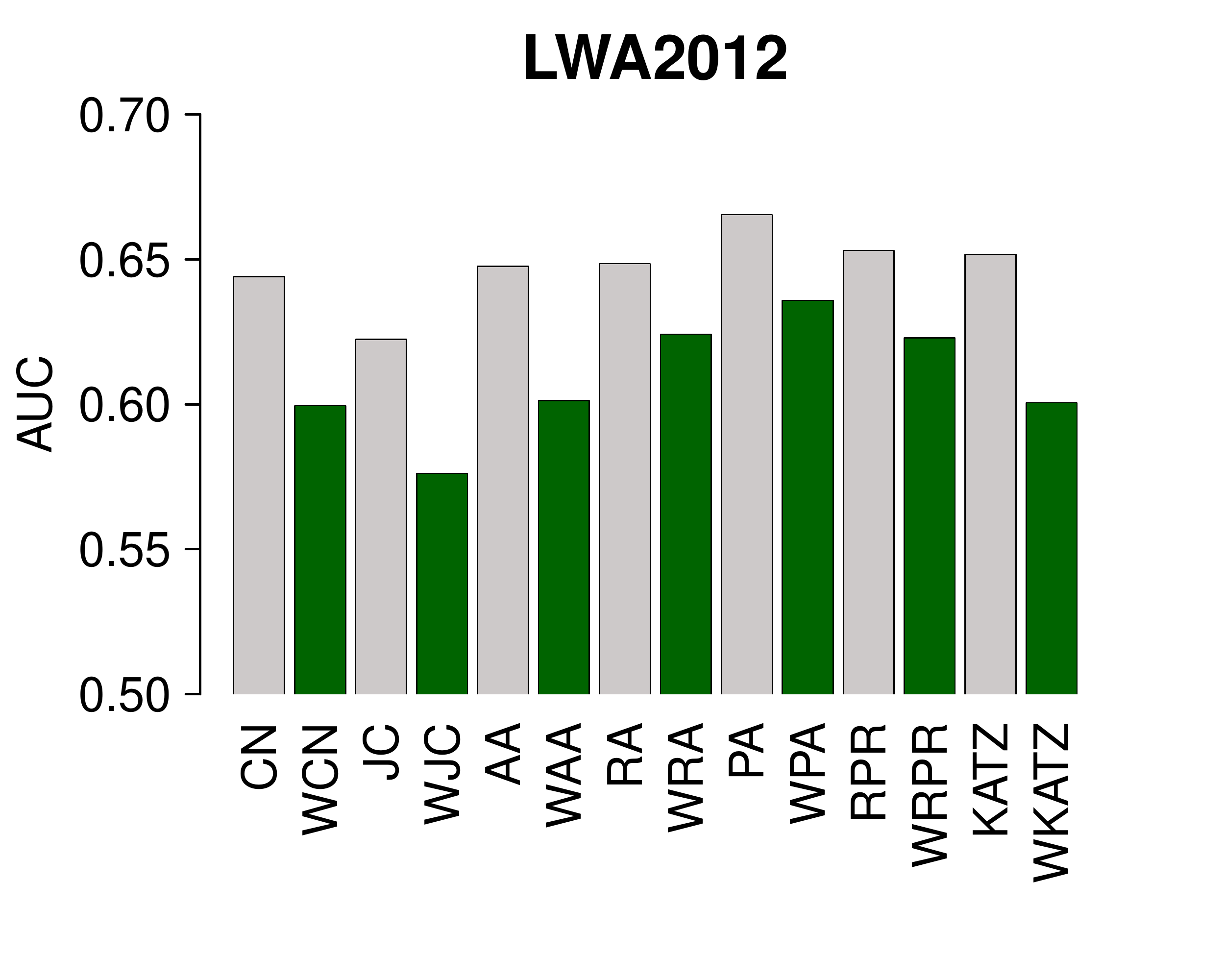}
  \caption{AUC values for new link prediction for each network proximity measure.}
  \label{fig:AUC_NEWLINKS_T0}
\end{center}
\end{figure}

In this subsection, we evaluate the quality of several link prediction measures (see Table \ref{tab:networkprosimitymeasures}) to predict new links, 
\ie all links in $E^{>t}_\mathrm{core}\setminus E^{\leq t}$. 
In Figure \ref{fig:AUC_NEWLINKS_T0}, we present the predictor scores of the original network proximity measure as well as the weighted variants of these measures. 
Figure \ref{fig:AUC_NEWLINKS_T0} suggests, that the network structure helps to improve the prediction accuracy, because all measures outperform the random predictor (the AUC value of a random predictor is $0.5$). This also means that in a human contact network the network topology contains useful information for the prediction
 of new links. This result is not surprising, since it confirms the results of \cite{Kleinberg2003} and \cite{Wang2011}. Here the authors analyzed the predictive power of 
 proximity network measures in a co-authorship network and a mobile phone caller network. For the HT 2011 and LWA 2010 datasets the weighted variants of 
 \textit{Resource Allocation} and \textit{weighted Adamic Adar} performed best. 
  In Figure\ref{fig:AUC_NEWLINKS_T0} we further compared the AUC values of the original and the weighted versions of the proximity measures: Considering the neighborhood-based network proximity measures we observe that the weighted variants always achieve better results than the unweighted versions at the LWA 2010 and HT 2011 dataset. However this observation does not hold on the LWA 2012 dataset: The network statistics in Table~\ref{tab:datasetstatistics} show, that for higher average contact lengths $\bigl(\mathit{ACL}($HT 2011$)< \ $$\mathit{ACL}($LWA 2010$)< \ $$\mathit{ACL}($LWA 2012$)\bigr)$ there is a tendency that unweighted variants tend to perform better compared to the weighted measures. We observe this tendency as an indicator of a certain bias of very strong links at the first day for the prediction of \emph{new} contacts. 

\begin{figure}[!htb]
   \includegraphics[width=.475\columnwidth]{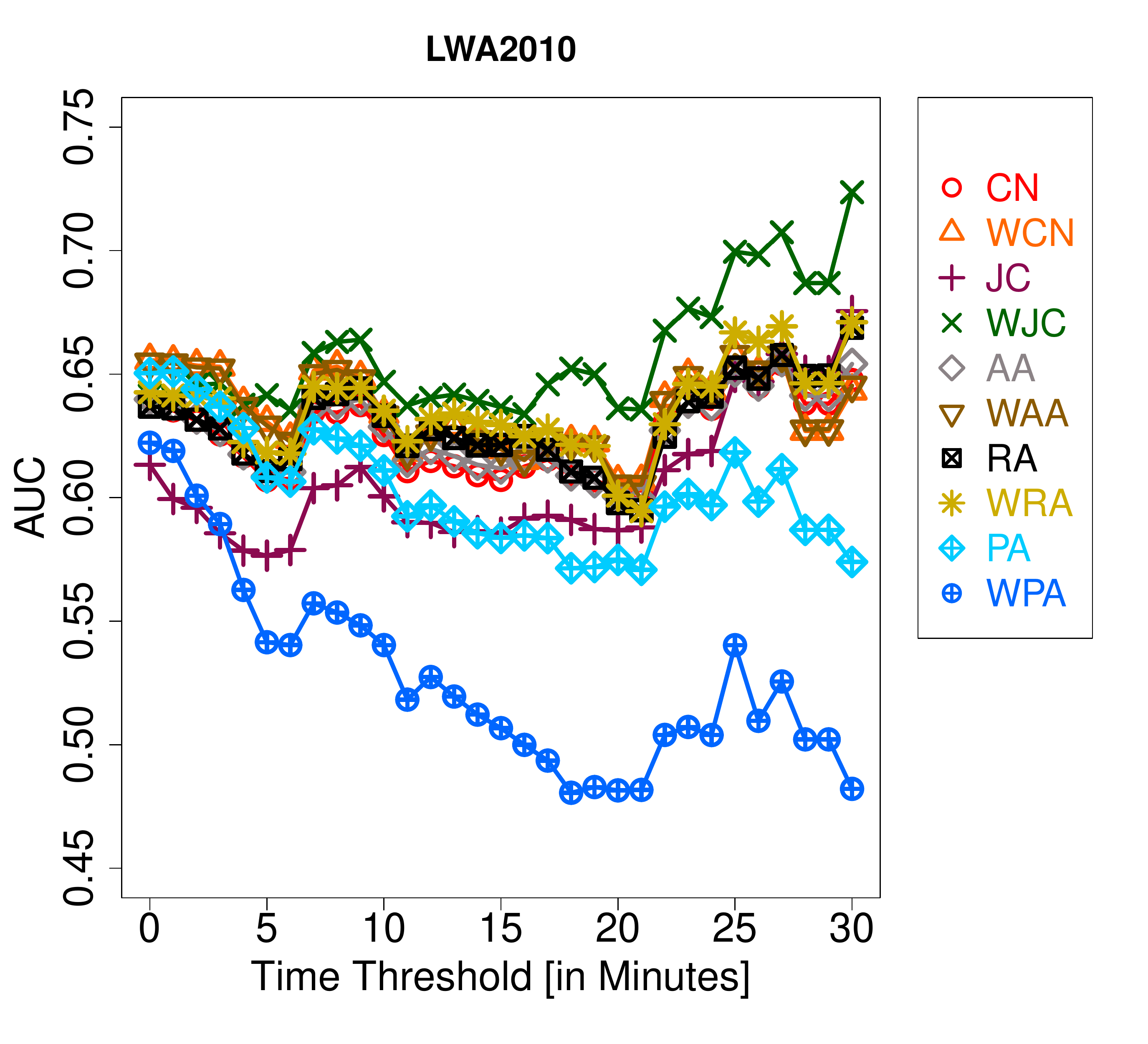}
      \includegraphics[width=.475\columnwidth]{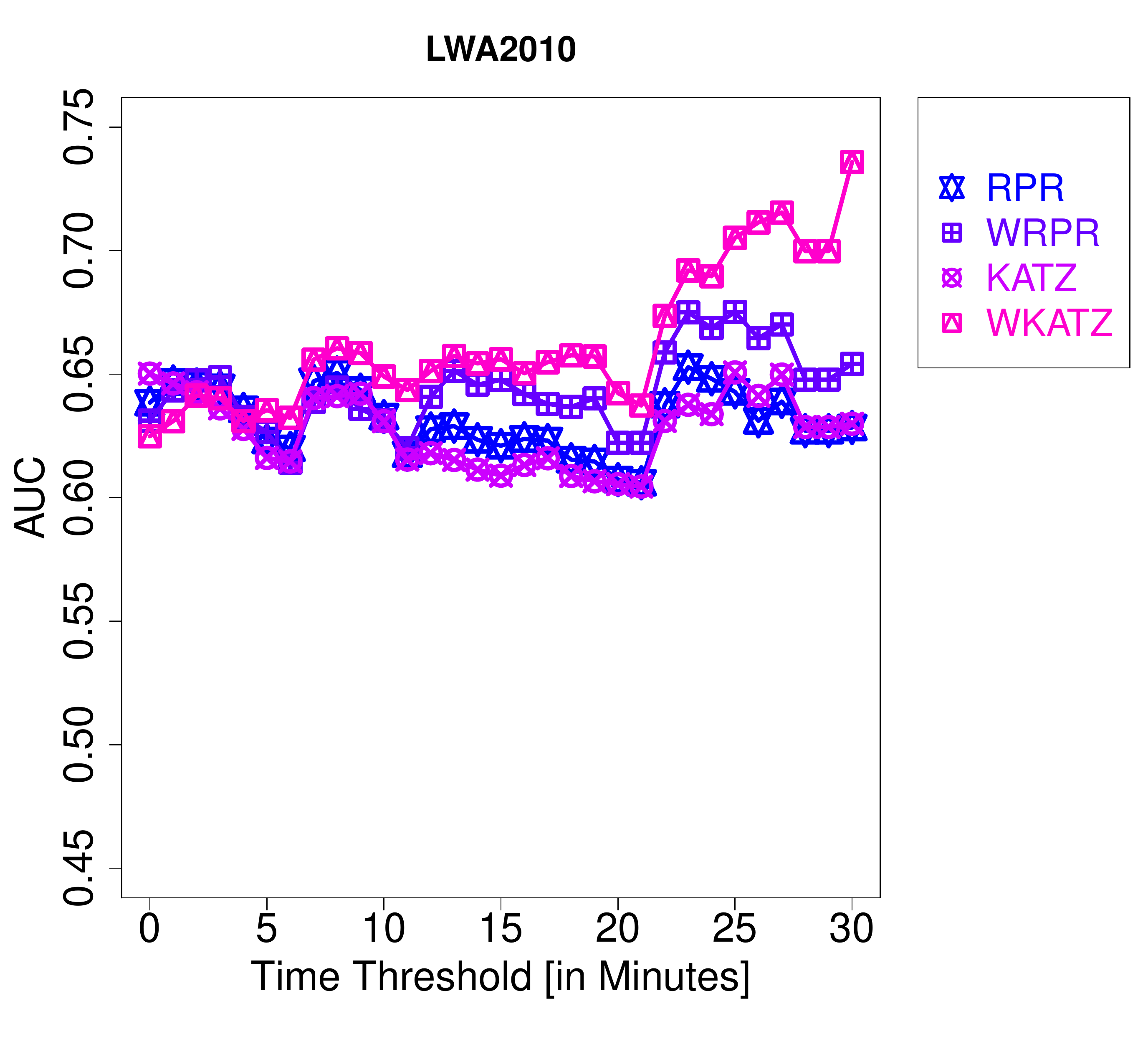}
      \includegraphics[width=.475\columnwidth]{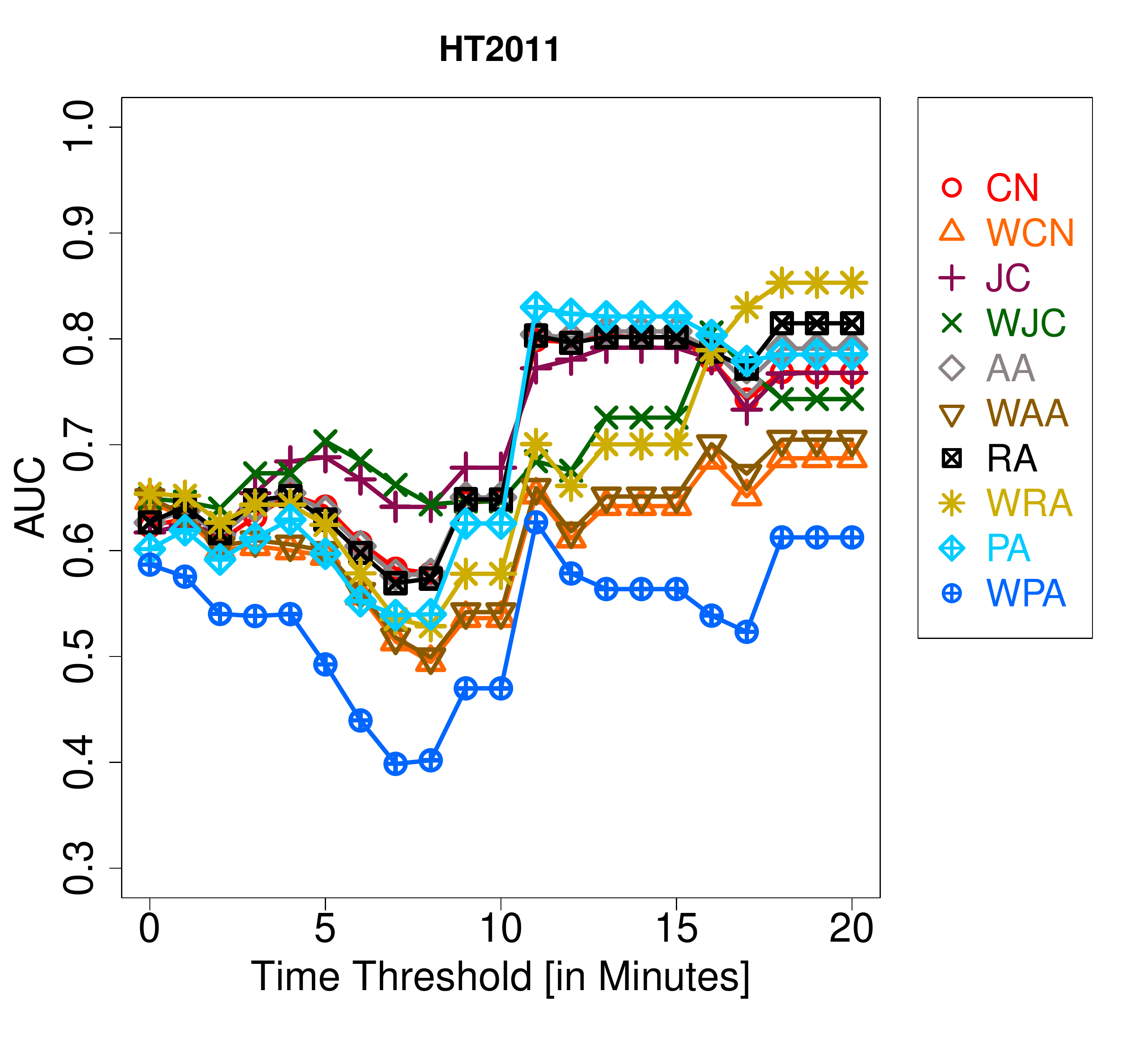}
      \includegraphics[width=.475\columnwidth]{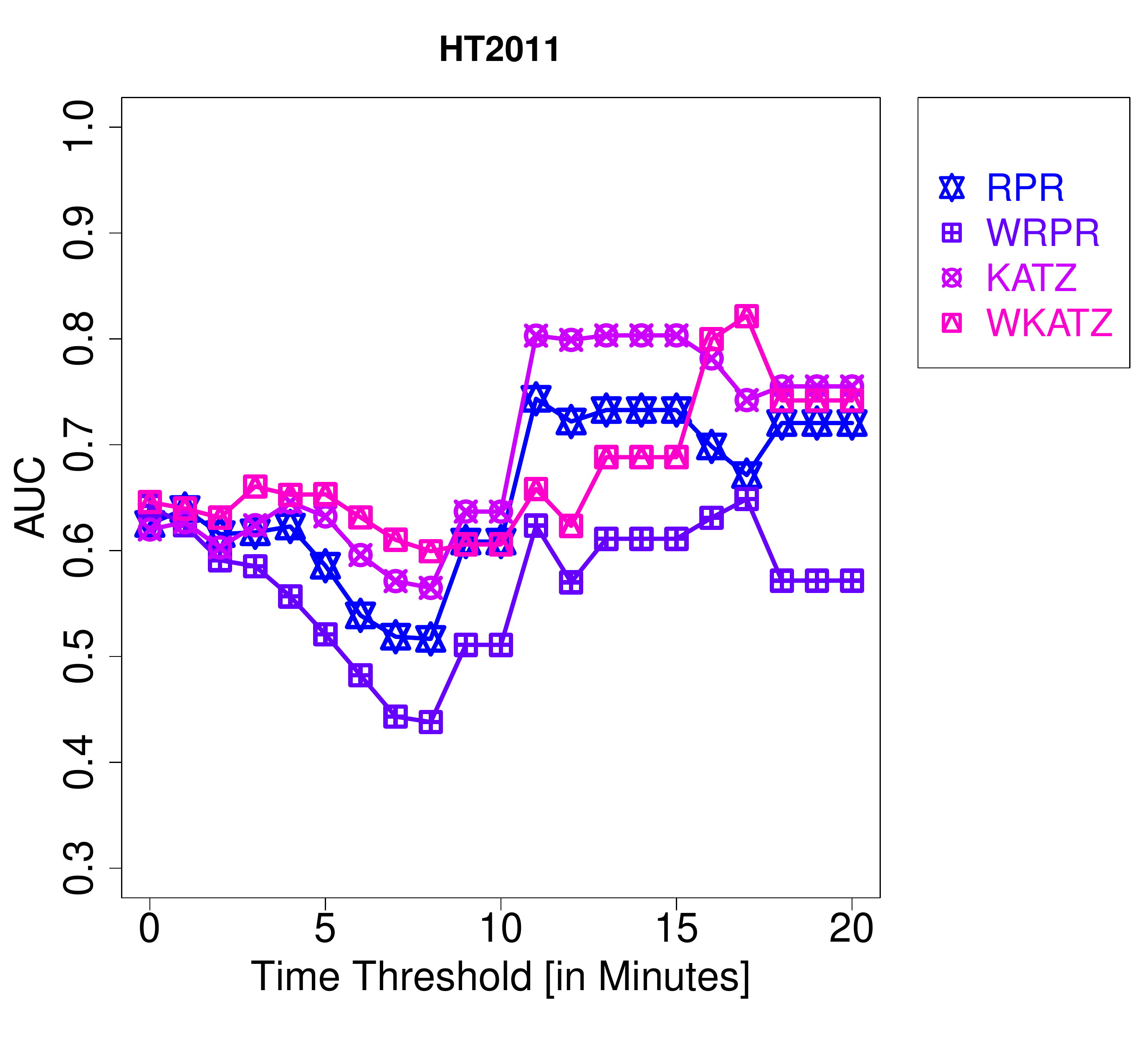}
         \includegraphics[width=.475\columnwidth]{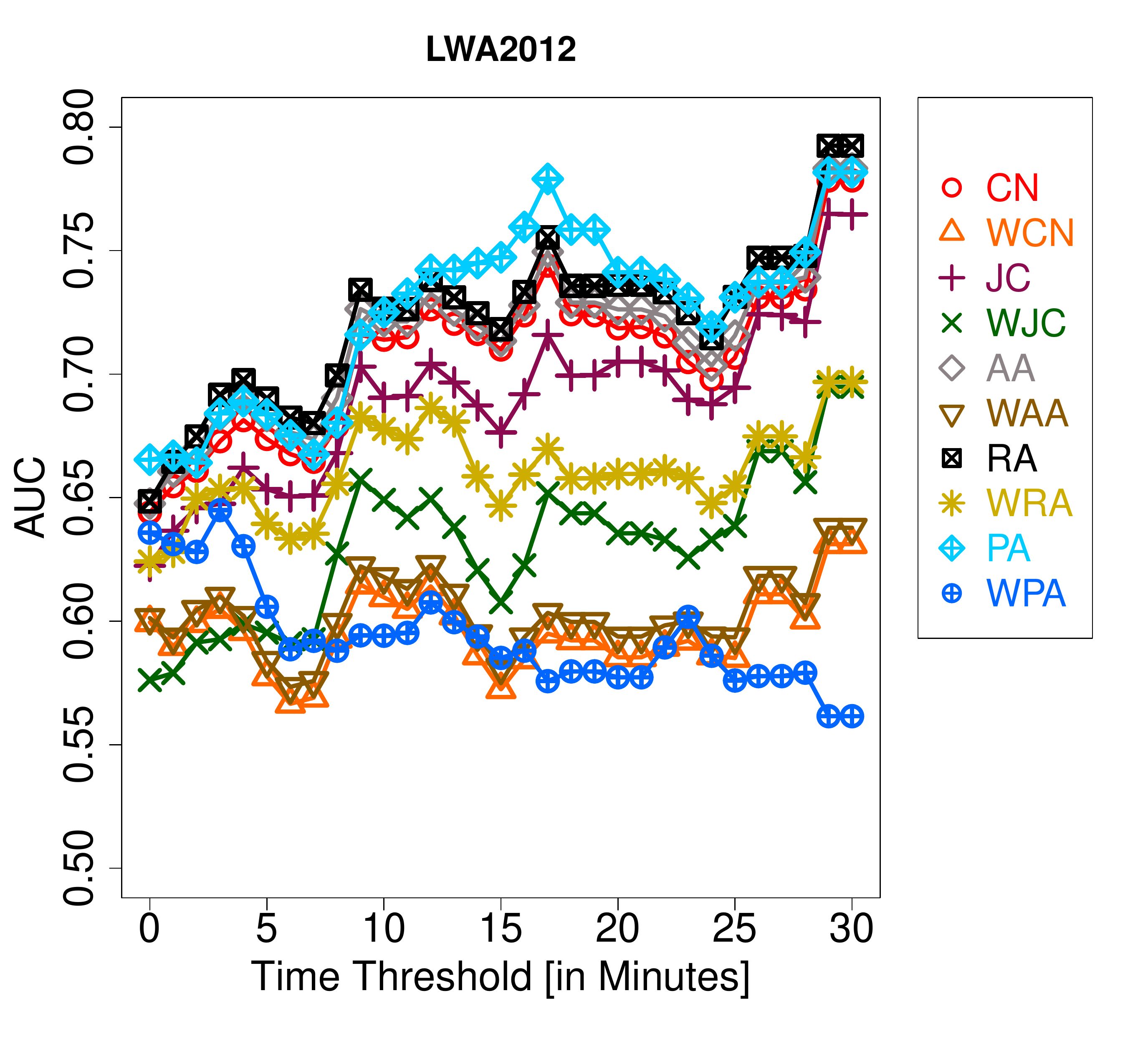}
      \includegraphics[width=.475\columnwidth]{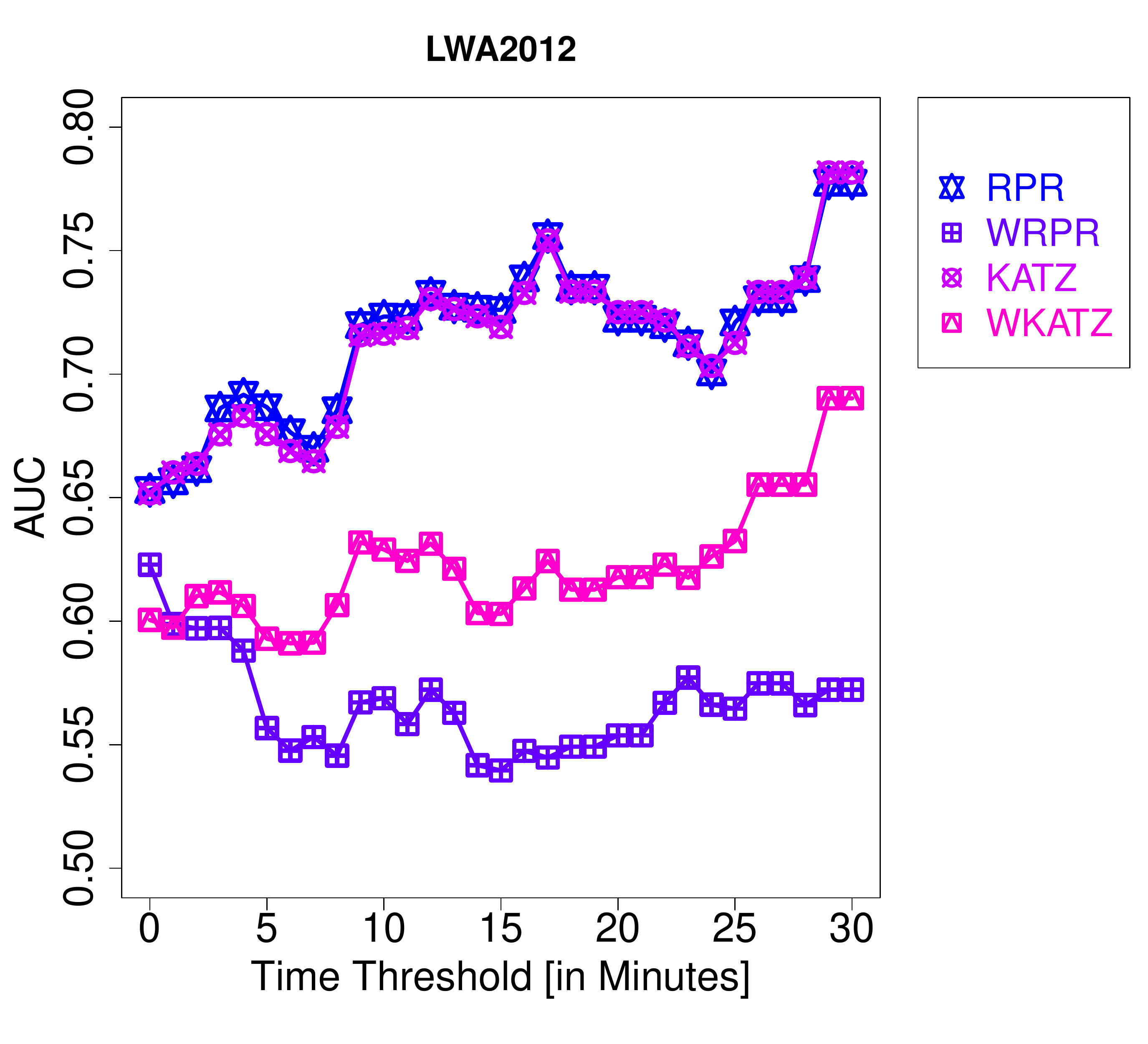}
  \caption{Threshold-based analysis of contact length and AUC-values for the prediction of new links, focussing on the ranking positions of longer face-to-face contacts, for the LWA 2010, the HT 2011 and the LWA 2012 conference.
The $x$-axis represents the minimum contact duration and the $y$-axis shows the AUC value for the prediction of new links with a contact duration at 
least this contact length.}
   \label{fig:PredictionContactLength}
\end{figure}

Figure \ref{fig:PredictionContactLength} shows the development of the AUC values for all neighborhood-based and path-based network proximity measure, when we focus more and more on longer contacts. 
 This means that we do not take into account contacts with contact length lower than a time threshold $t$ (value on the x-axis) and examine only
 the ranking positions of contacts
 greater than the time threshold $t$. In Figure~\ref{fig:PredictionContactLength} we see an interesting development. On all datasets we observe that longer face-to-face contacts tend to be placed higher in the ranking than shorter contacts. In addition we observe that the \emph{weighted Preferential Attachment} predictor performs very weak for stronger ties. Comparing the weighted and unweighted measures we could not find significant differences in prediction performance for stronger ties.


 \subsection{Influence Factors for the Prediction of Recurring Links}\label{sec:analysis:influence:recur}
 
In the recurring link prediction problem we want to predict whether a link between two participants $u$ and $v$ will recur or not. Unlike the \emph{new link} prediction problem~\cite{Kleinberg2003}, in the recurring link prediction problem we can also use the information about the already existing tie strength of the corresponding participants $u$ and $v$.

\begin{figure}[!htb]
 \begin{center}
      \includegraphics[width=0.4\columnwidth]{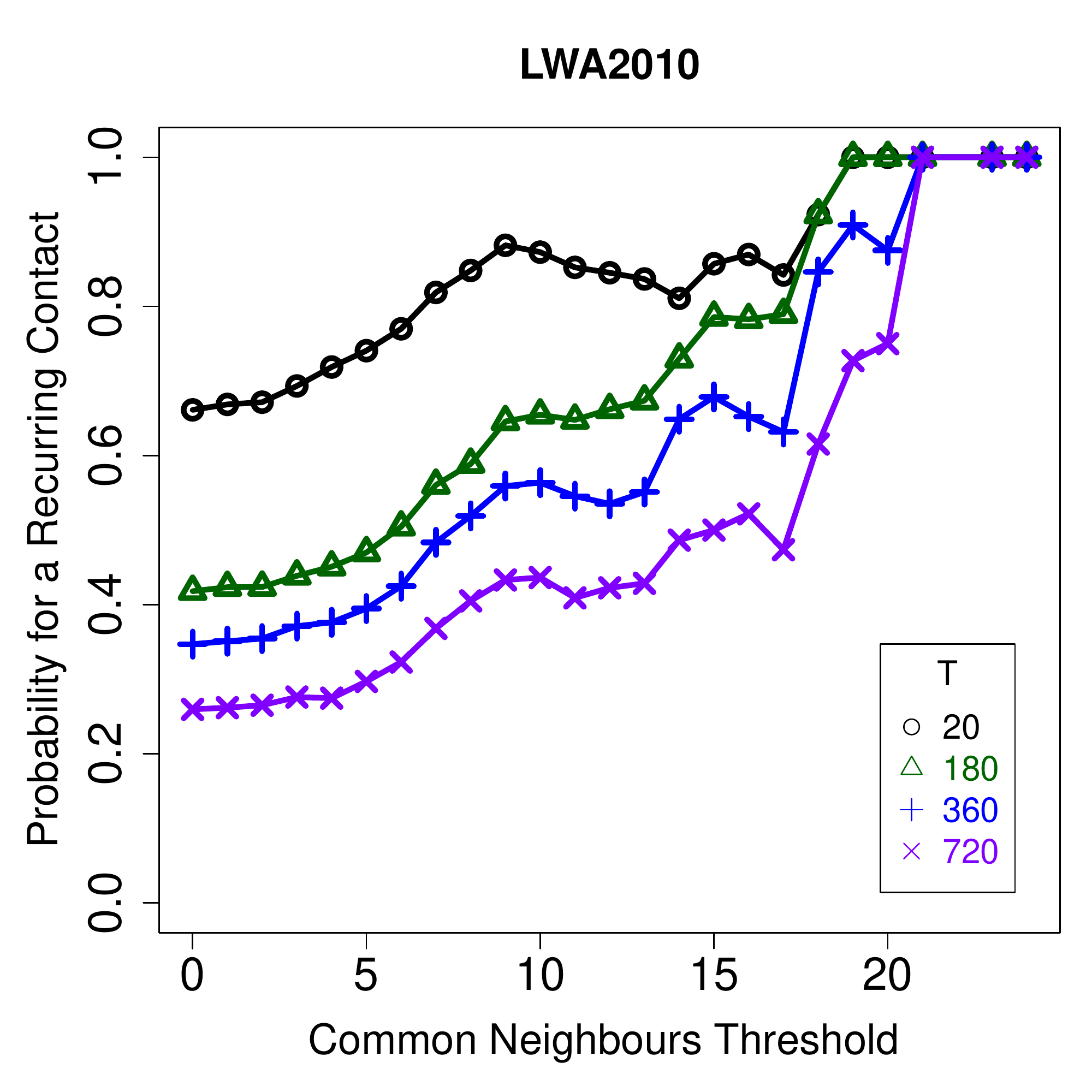}
      \includegraphics[width=0.4\columnwidth]{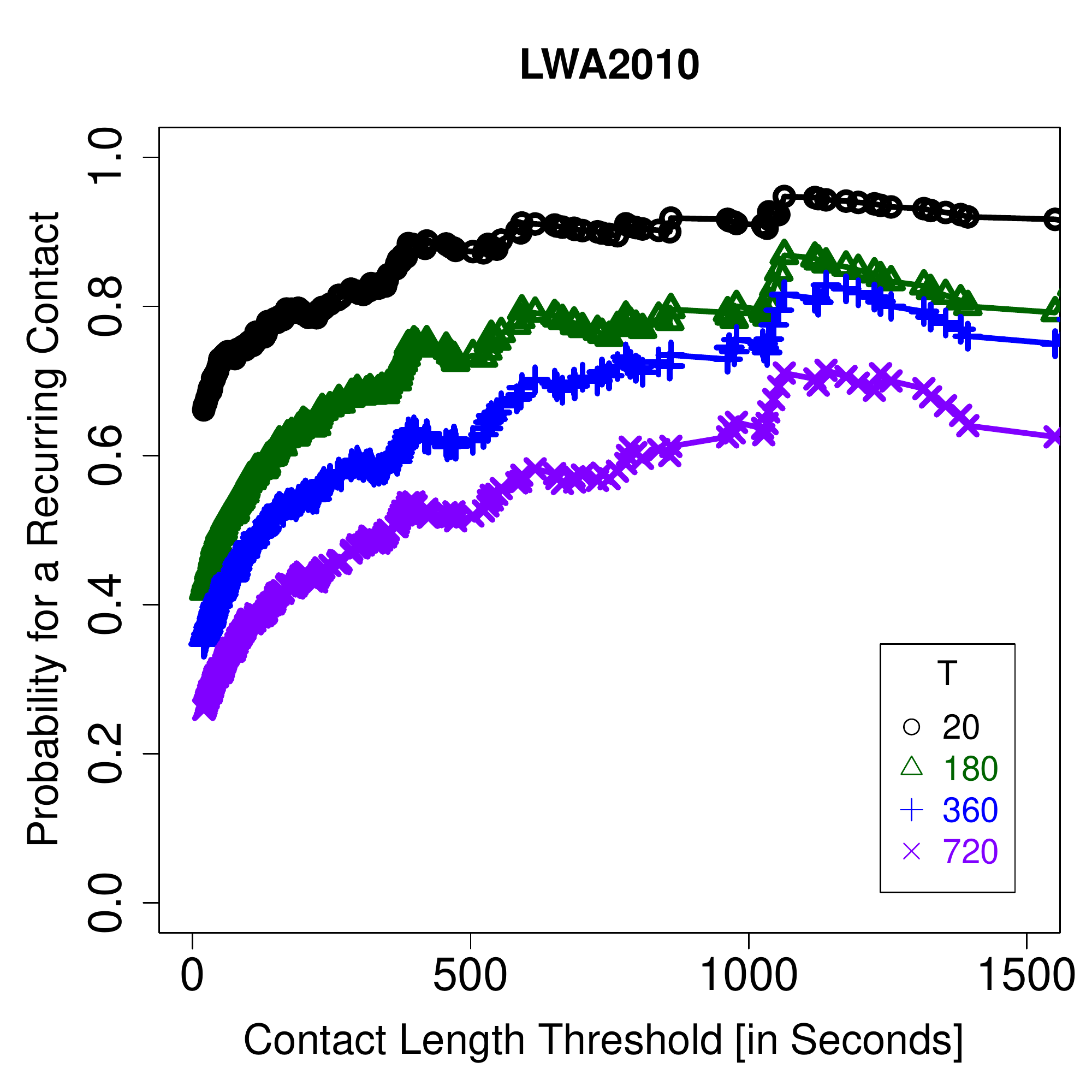}
 \end{center}
 \caption{Probability for a recurring link with strength $T$ as a function of common neighbours and tie strength. In both figures the $y$-axis shows the probability for a recurring link (with tie strength $T$), given at least: a) a specific number of common neighbors or b) a specific tie strength. The respective thresholds are defined by the $x$-axis.}
  \label{fig:ProbRecContact}
\end{figure}

Therefore, we analyze the influence of the number of common neighbors; in addition, we include the already existing tie strength on the recurrence of a link, complementing the analysis and questions already raised in Section~\ref{sec:analysis:basic}. In Figure \ref{fig:ProbRecContact}, we plot the probability for a recurring link with tie strength $T$ as a function of common neighbors and as function of the already existing tie strength. Given a face-to-face contact between two participants at the first day of the conference, we compute in this analysis whether a contact (with minimum contact duration $T$) recurs or not on the second or third day of the conference, depending on the number of common neighbors and existing tie strength of the first day. 
 We observe, that the probability increases almost linearly the higher the number of common neighbors and the higher the already existing tie strengths are.

\subsection{Predictability of Recurring Links in Face-to-Face Proximity Networks}


In this section we evaluate and compare the quality of network-based and path-based network proximity measures to predict recurring links. Furthermore we use the \emph{current tie strength} between two participants as predictor. For our prediction analysis, we compute these predictor scores, based on the face-to-face proximity network of the first day of the conference. Then we use these predictor scores to analyze its prediction quality with respect to whether a link will recur or not towards the end of the conference.

\begin{figure}
\begin{center}
      \includegraphics[width=0.4\columnwidth]{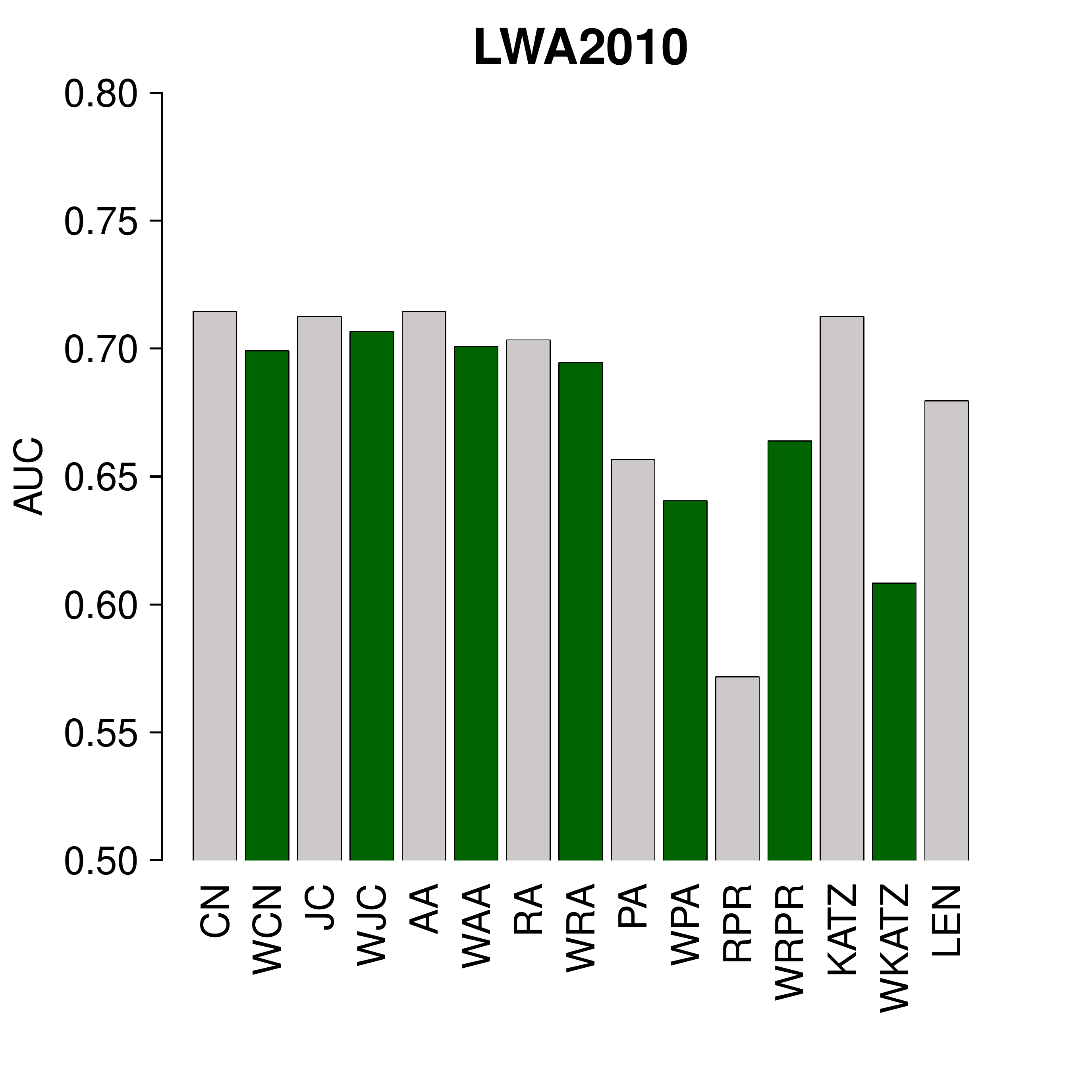}
      \includegraphics[width=0.4\columnwidth]{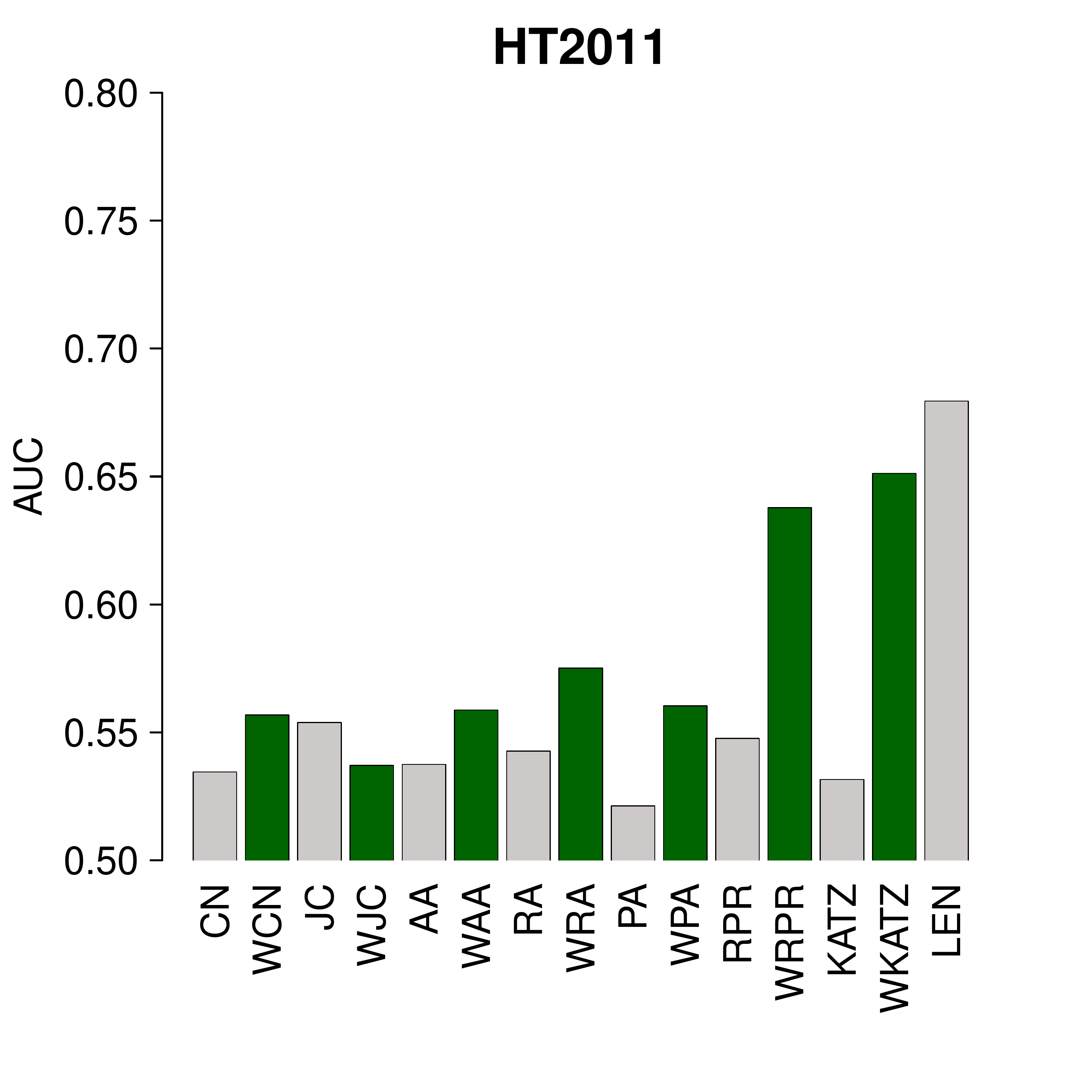}
      \includegraphics[width=0.4\columnwidth]{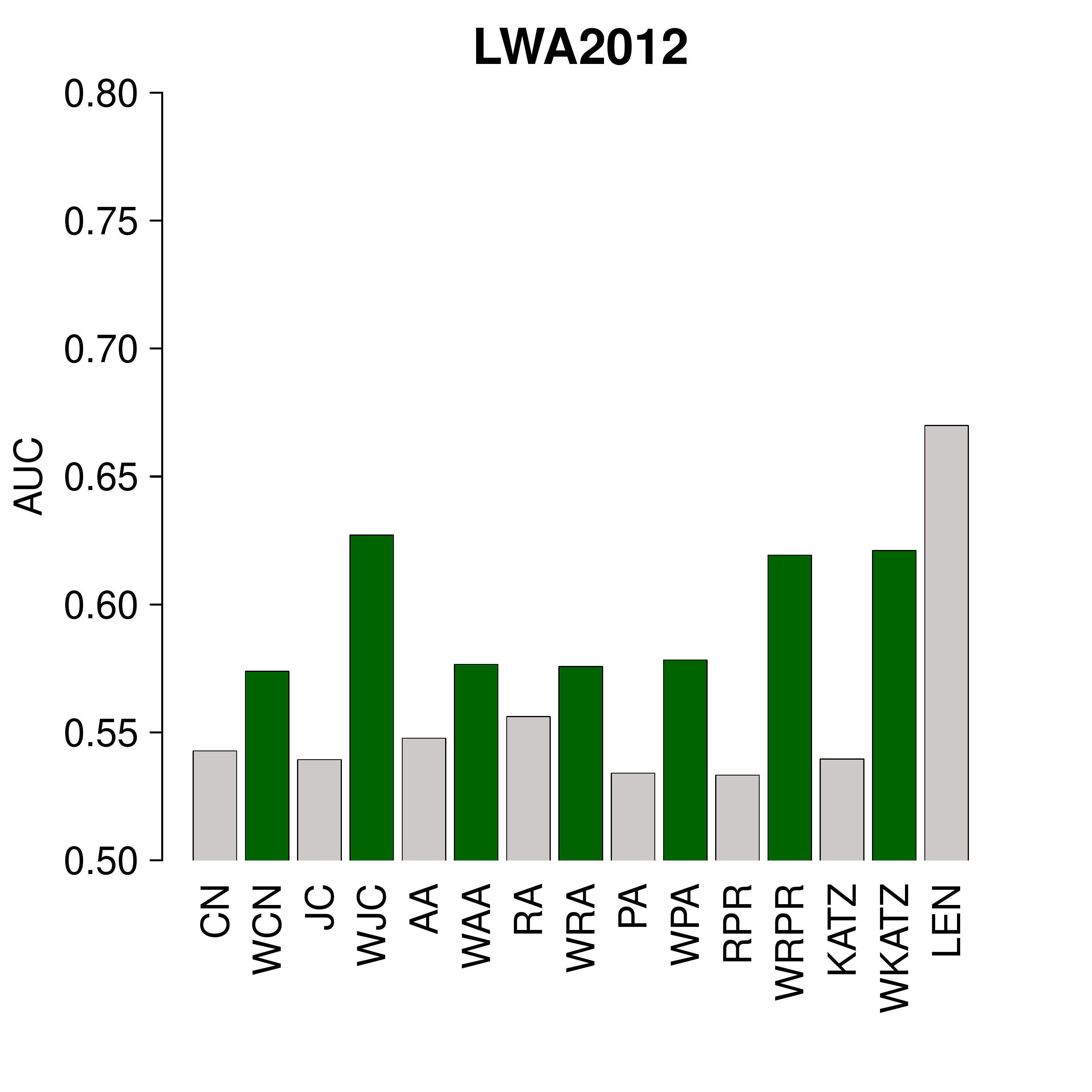}
  \caption{AUC values for recurring link prediction for each network proximity measure. LEN here indicates the \emph{current tie strength} as predictor.}
  \label{fig:AUC_RECLINKS_T0}
  \end{center}
\end{figure}

 In Figure \ref{fig:AUC_RECLINKS_T0}, we plot the AUC-values for all network proximity measures measures and the \emph{current tie strength} (LEN). First, we observe that  the network structure helps to improve the prediction accuracy, because all predictors outperform the random predictor. Furthermore we notice that the first day's tie strength performs very well as predictor on all datasets. With respect to the HT 2011 and LWA 2012 dataset we see that path-based network proximity measures perform better than measures based on the nodes' neighborhood. However this result does not hold on the LWA 2010 dataset.
 
\begin{figure*}[!htb]
     \includegraphics[width=0.495\columnwidth]{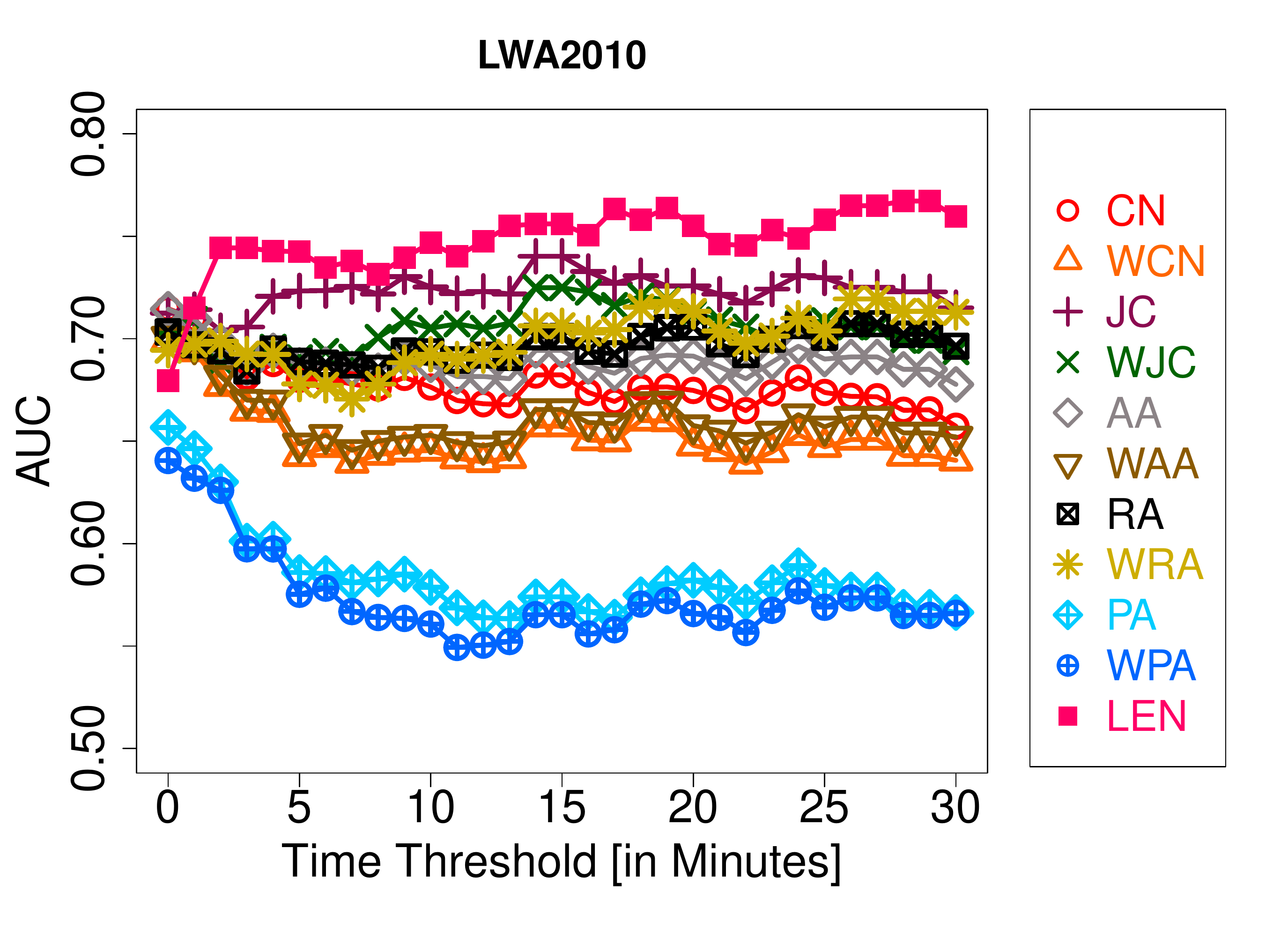}
     \includegraphics[width=0.495\columnwidth]{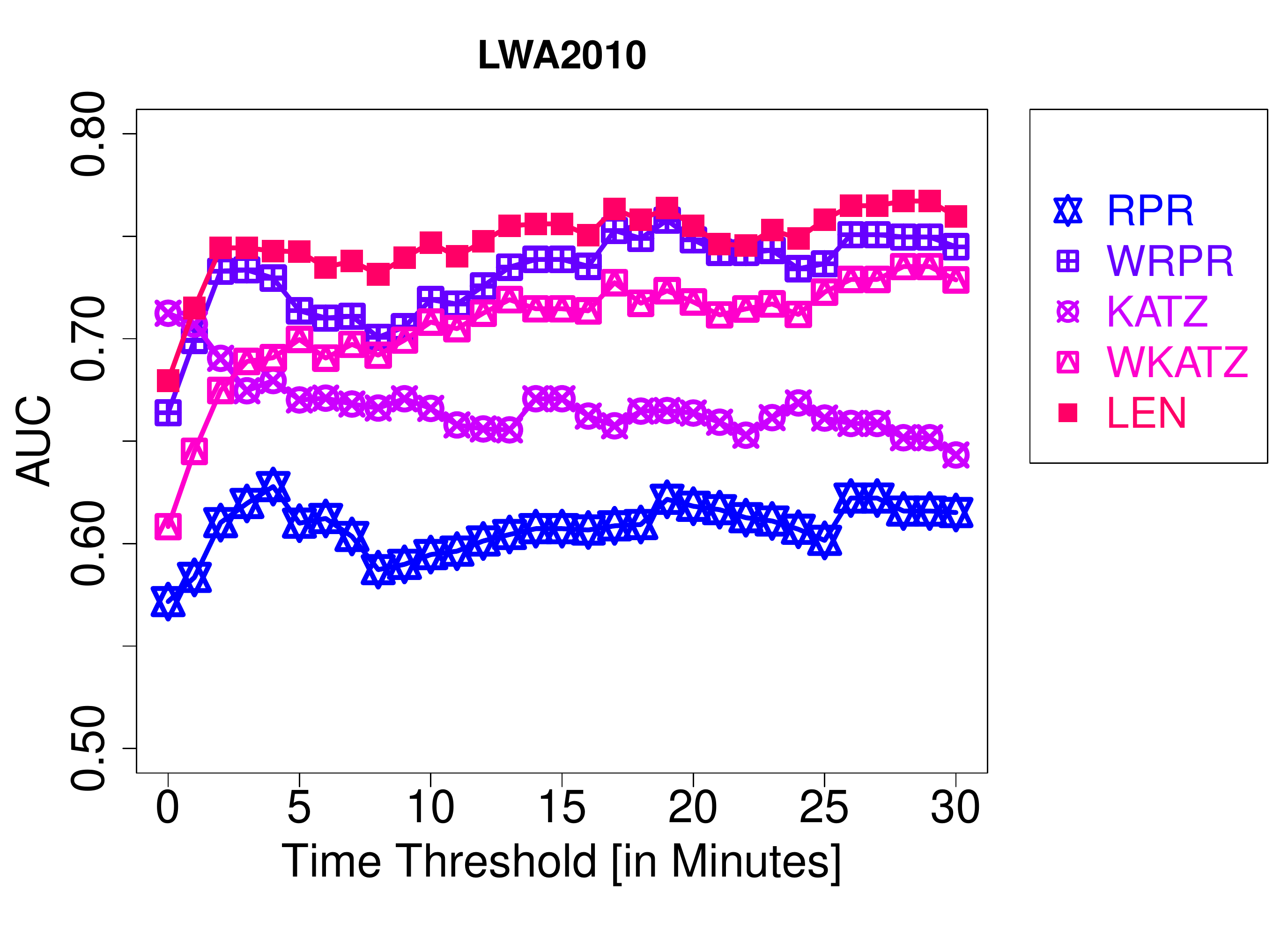}
     \includegraphics[width=0.495\columnwidth]{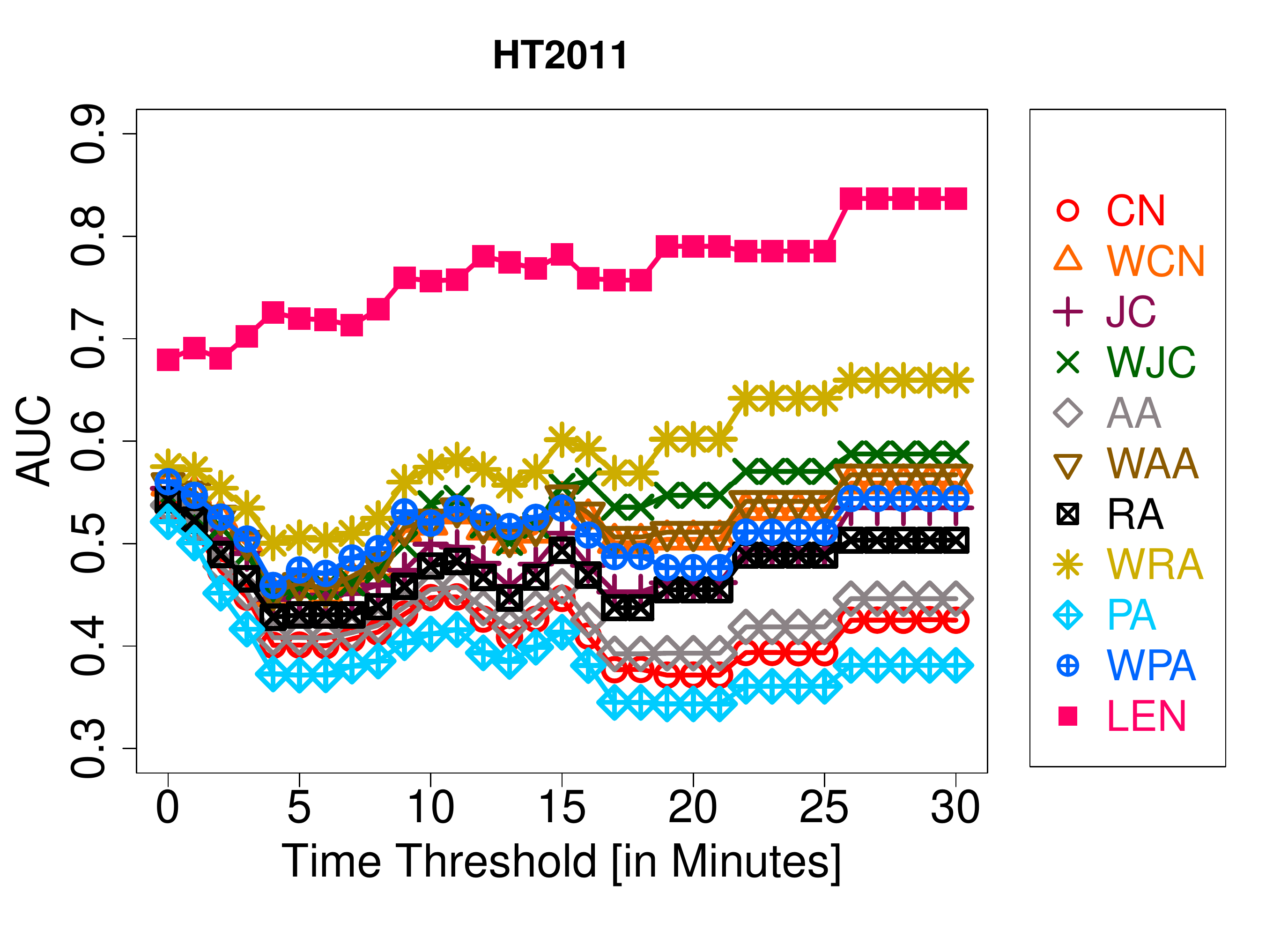}
     \includegraphics[width=0.495\columnwidth]{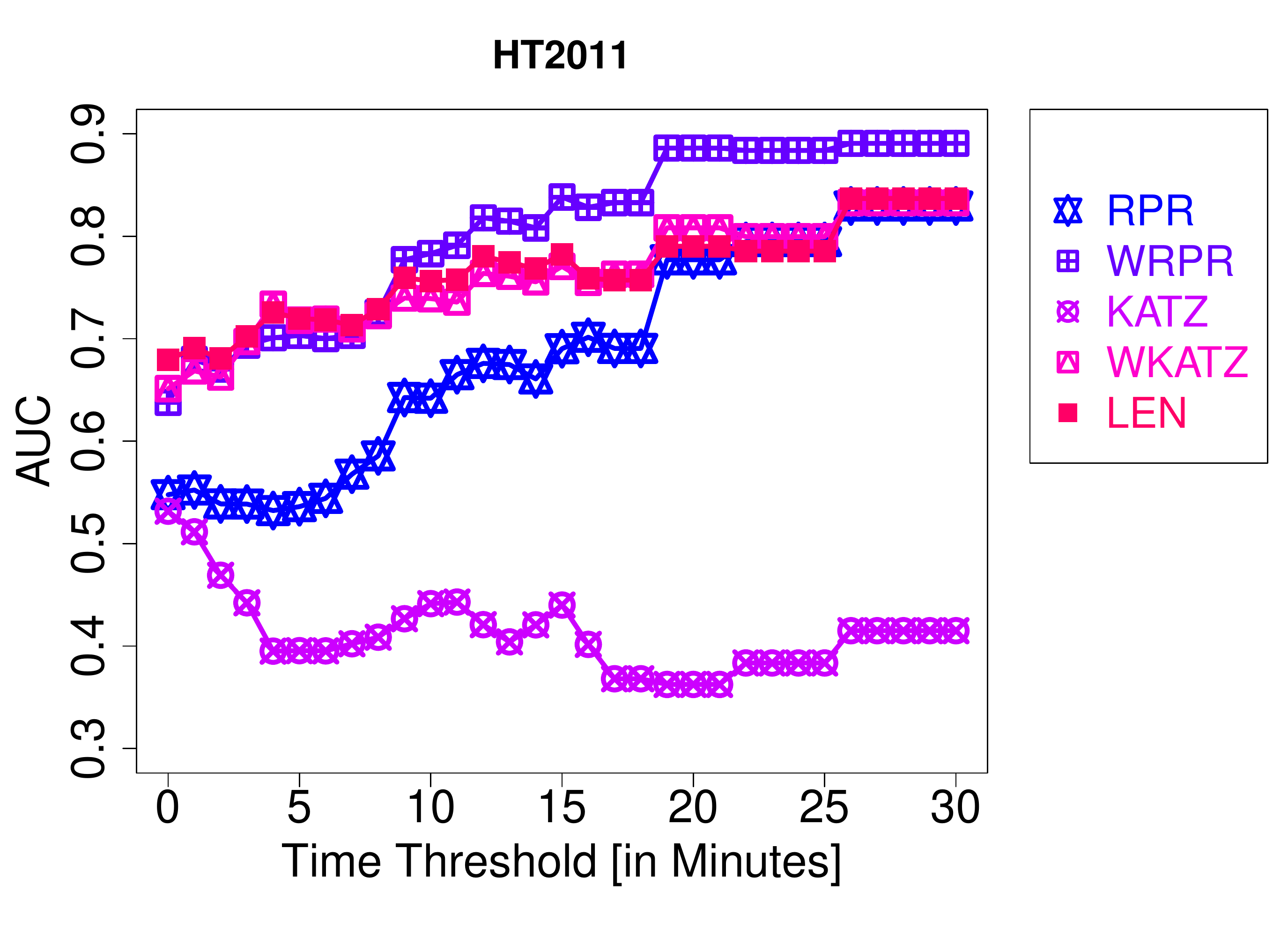}
     \includegraphics[width=0.495\columnwidth]{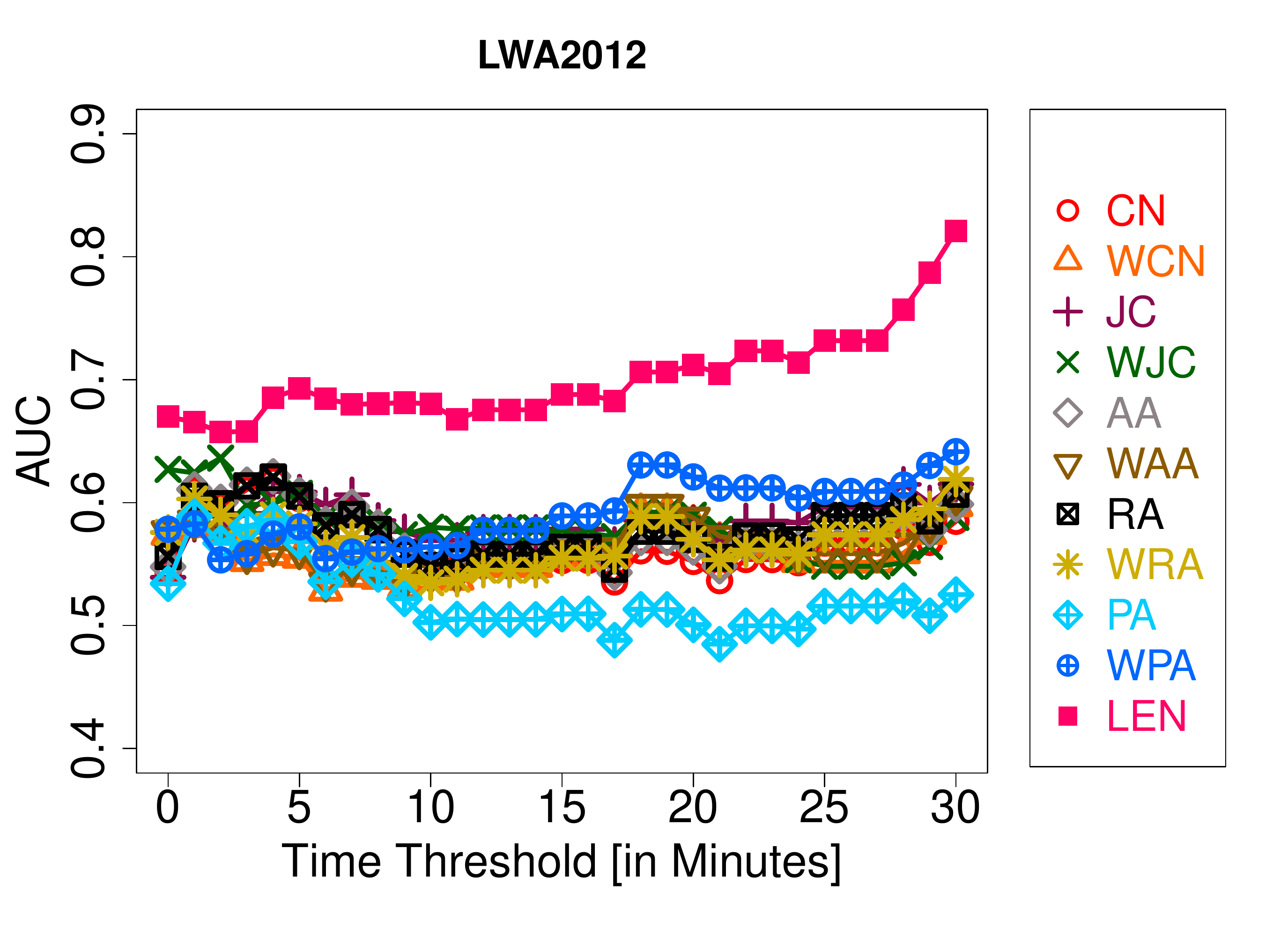}
     \includegraphics[width=0.495\columnwidth]{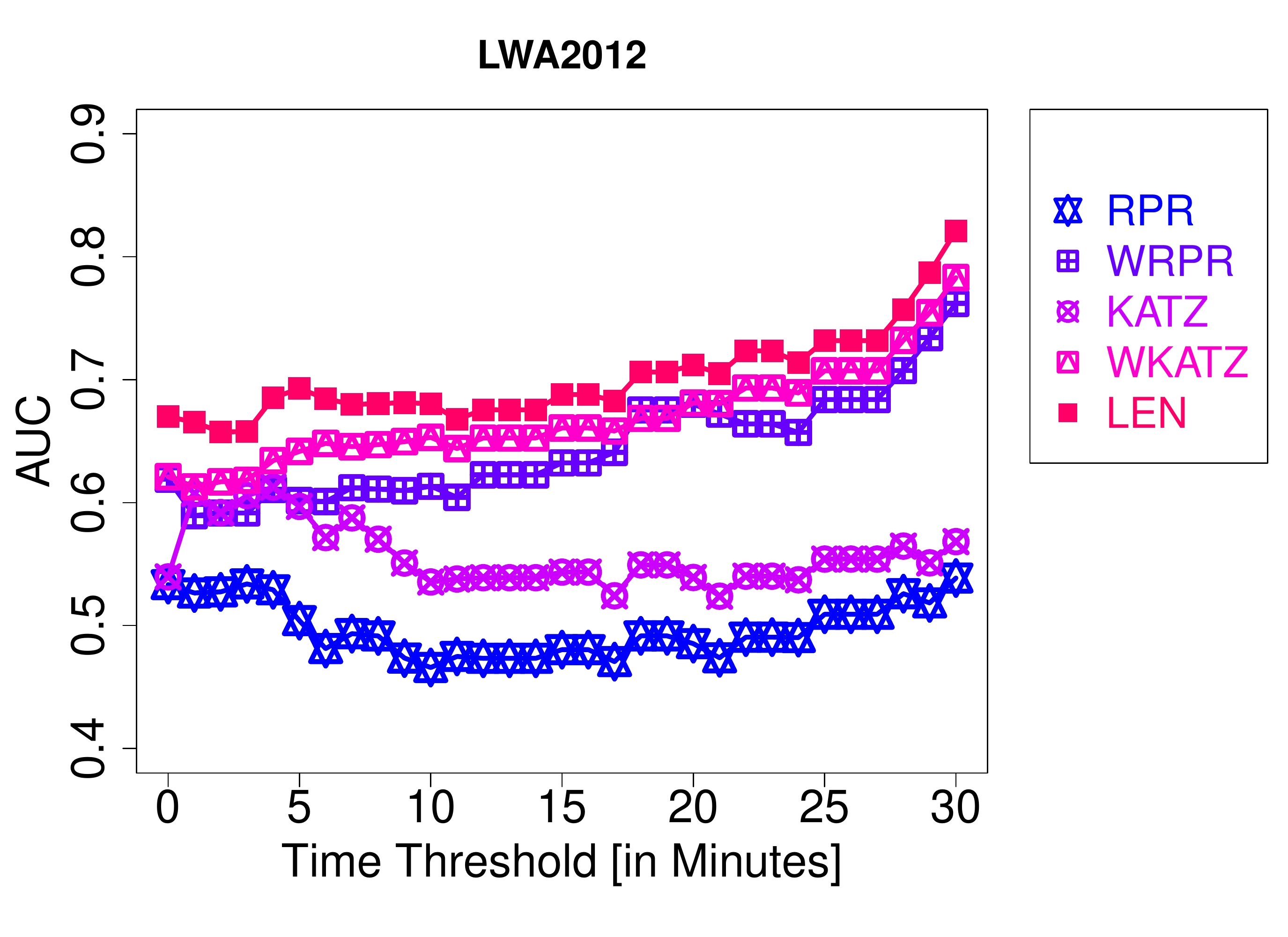}
  \caption{AUC values for recurring link prediction for different time thresholds. The $y$ represents the AUC value for the given time threshold $t$ (defined by the $x$-axis). We note here that we only consider future links with tie strength $>=T$ for the prediction task. LEN here indicates the \emph{current tie strength} as predictor.}
  \label{fig:AUC_RECLINKS_T0_30}
\end{figure*}

 In Figure~\ref{fig:AUC_RECLINKS_T0_30}, we focus more and more on longer face-to-face contacts for the link prediction task. This means  that we only consider face-to-face contacts longer than a given time threshold $T$. In Figure~\ref{fig:AUC_RECLINKS_T0_30}, this time threshold $T$ is defined by the $x$-axis. Considering longer contacts,  we observe that weighted path-based measures clearly outperform network proximity measures based on the nodes' neighborhood. Furthermore, the figure shows that the weighted variants of the path-based measures perform much better than the unweighted variants. In addition we notice that (also for longer contacts) using the first day's tie strength as predictor performs very well on all datasets. Except for the HT 2011 dataset, this predictor performs best. This is surprising, because we expected that the path-based measures would significantly outperform all other measures. Apparently, the combination of information of the node's neighbourhood with the first day's tie strength is boosting the performance. Considering the neighbourhood-based measures, we see that the unweighted \emph{Preferential Attachment} predictor performs very weak on all datasets.

\subsection{The Role of Weak Ties for the Prediction of Recurring Links}

\begin{figure}[!htb]      
      \begin{center}
      \includegraphics[width=0.45\columnwidth]{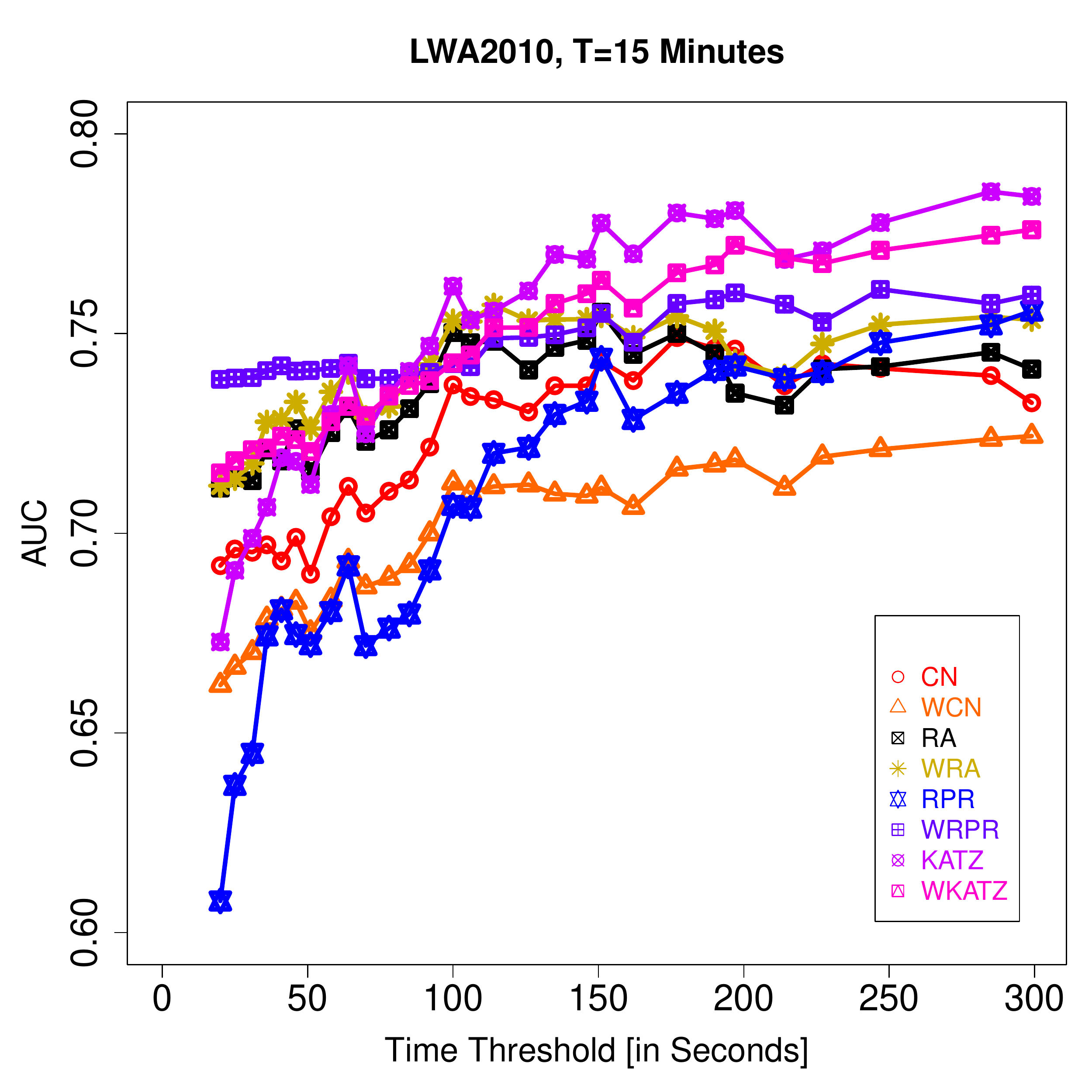}
\includegraphics[width=0.45\columnwidth]{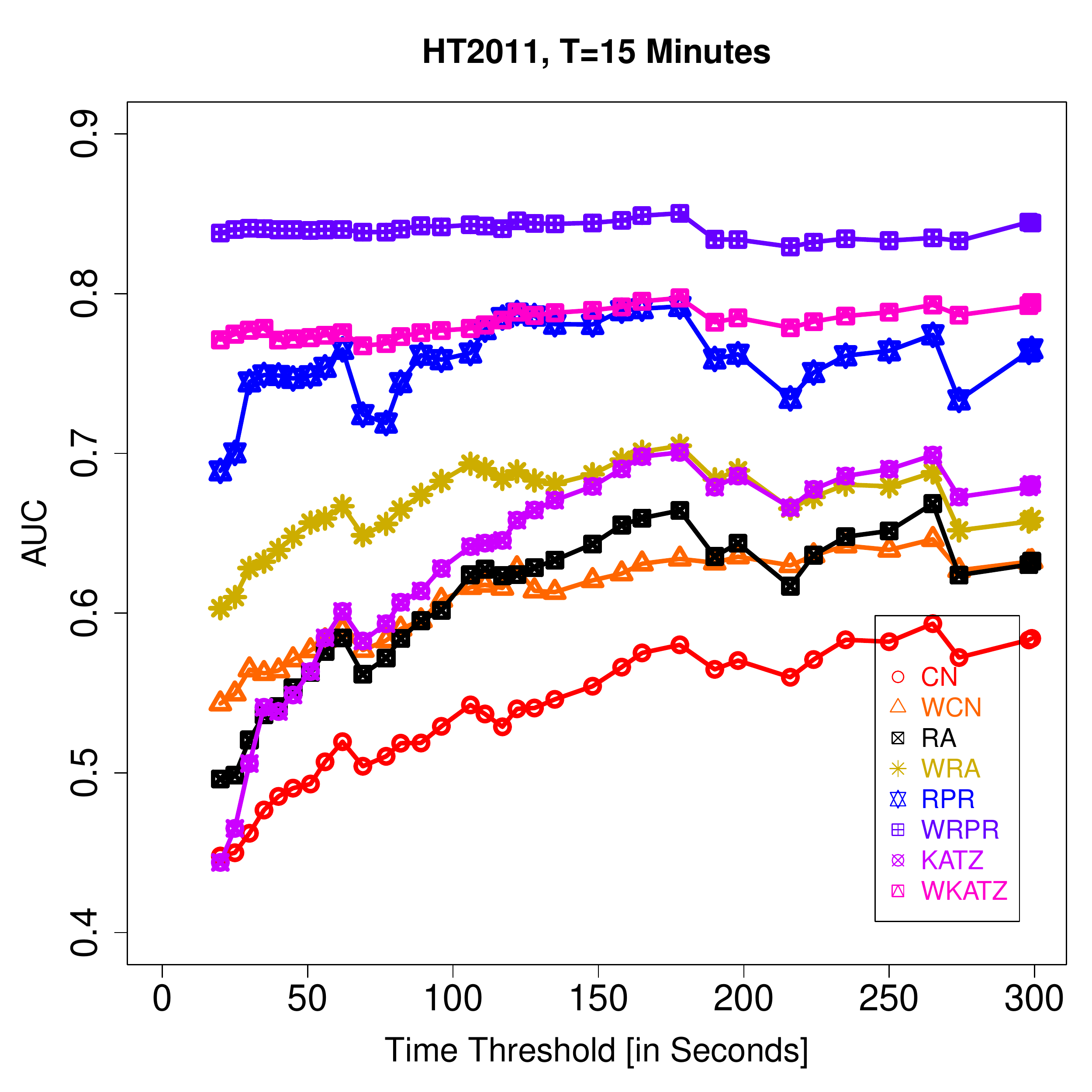}
\includegraphics[width=0.45\columnwidth]{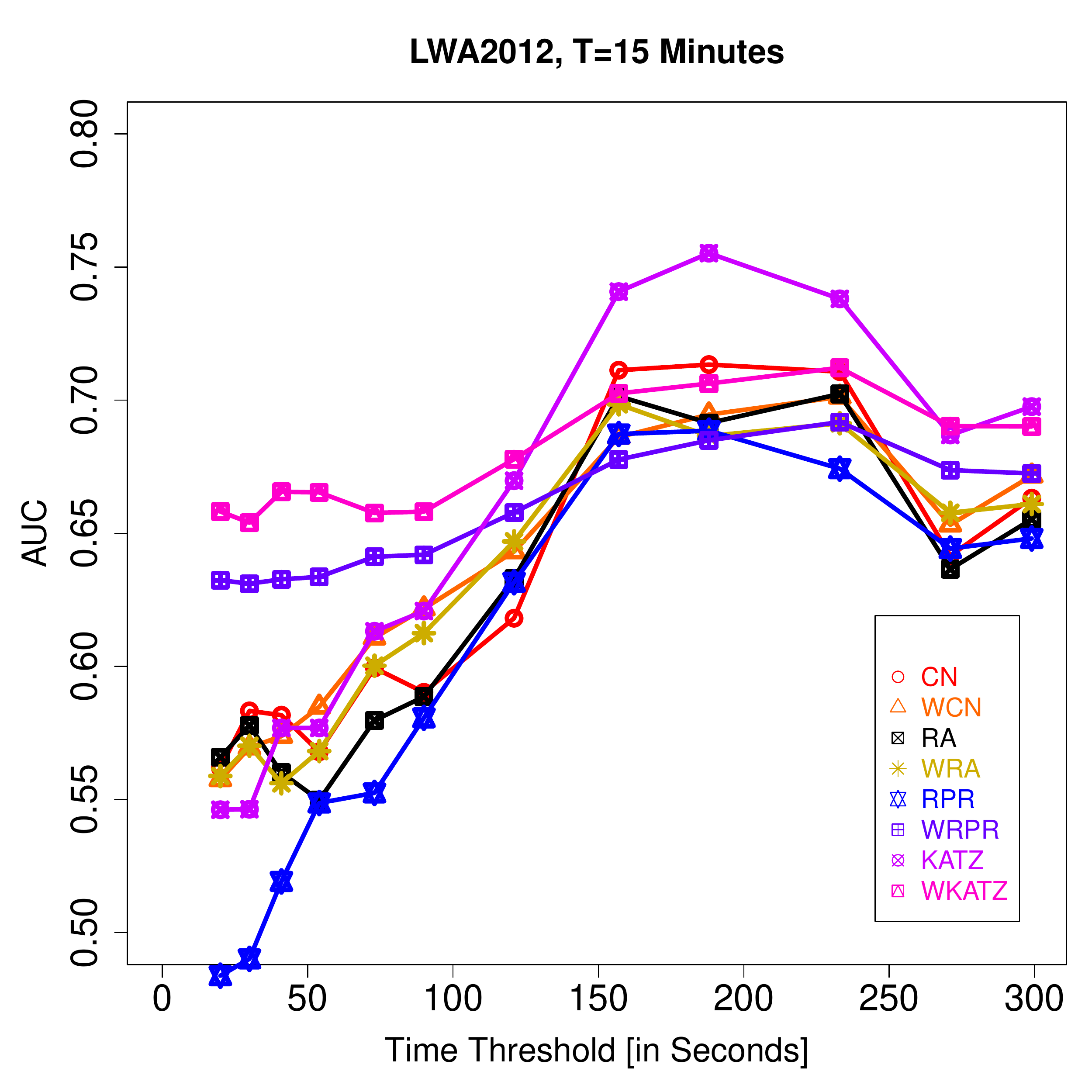}
  \caption{AUC values for several network proximity measures for predicting recurring links, when we remove all links below a certain time threshold $T$ is given on the $x$-axis.}
  \label{fig:AUC_RECLINKS_RM}
\end{center}

\end{figure}

We also analyse the role of weak ties for our prediction scenario. Exemplarily we focus here on the prediction of stronger links with a time threshold of $15$ minutes, but the results are very similar for other time thresholds. For the analysis, we compute the AUC value for several network proximity measures, using the face-to-face contact networks, where all links have been removed that fall below a given time threshold $T$. In Figure~\ref{fig:AUC_RECLINKS_RM}, this threshold $T$ is defined by the $x$-axis. We observe that the removal of weak links increases the prediction accuracy of most network proximity measures. Especially on the LWA 2010 and LWA 2012 datasets the AUC value for the unweighted rooted PageRank increases for more than $15 \%$ AUC, when we remove all links weaker than 200 seconds. Considering this threshold, we can also observe an increase of AUC for all weighted and unweighted neighbourhood-based network proximity measures.

 For the \emph{weighted rooted PageRank} predictor, we observe the interesting trend that the removal of weak ties seems to have less influence concerning the prediction accuracy. This stability can be explained by the fact that the \emph{weighted rooted PageRank} also uses the information of the first day's tie strength. Except for the LWA 2010 dataset, this result is also true for the weighted Katz predictor.

%% file: conclusions.tex
\section{Conclusions}\label{sec:conclusion}
In this paper, we considered the predictability of human face-to-face contacts and presented an analysis of influence factors for link prediction in such human contact networks. Specifically, we considered the standard problem of predicting new links, and extended it to the analysis of recurring links. We compared the performance of path-based and neighbourhood-based network proximity measures for predicting new and recurring links. Considering recurring links we also studied the \emph{current tie strength} as predictor. We observed that stronger links are better predictable for the new and recurring link prediction problem. Especially the weighted variants of the path-based network proximity measures perform much better in the prediction of recurring links than neighbourhood-based network proximity measures. The results also show, that the current tie strength performs better than the path-based measures on two of the three datasets. This is surprising, because path-based measures combine information from the current tie strength and the nodes' neighbourhood. Furthermore, we studied the predictability of recurring links, when weak links are removed from the network. We observed that removing links with weight (contact length) smaller than 200 seconds increases the AUC-values for most network proximity measures. Furthermore, we considered (and adapted) different network proximity measures for the prediction, and took descriptive properties of human participants into account.
These insights are a first step onto predictability applications for human contact networks, \eg for improving recommender systems.
  
For future work, we aim to leverage these analysis results in order to embed the indicators, patterns, and influence factors into more advanced prediction models in the context of human contact networks. Furthermore, we plan to extend the analysis towards more dynamic 
approaches including movement and location-based events for improving the prediction further.